\documentclass[a4paper]{jpconf}
\usepackage{graphicx}
\usepackage{amsmath}
\usepackage{xcolor}

\usepackage{amssymb}
\usepackage{amsfonts}

\newcommand{\neff}{n_{\text{eff}}}

\newcommand{\nefft}{\tilde{n}_{\text{eff}}}
\newcommand{\kappsiS}{\kappa_{\scriptscriptstyle \psi_{\!s}}}
\newcommand{\psubsup}[2]{p_{\scriptscriptstyle \!#1}^{\scriptscriptstyle (\!#2\!)}}
\newcommand{\Qsub}[1]{Q_{ #1}}
\newcommand{\nsub}[1]{n_{ #1}}
\newcommand{\psub}[1]{p_{ #1}}
\newcommand{\kapsub}[1]{\kappa_{ #1}}
\newcommand{\Chatsub}[1]{\hat{C}_{ #1}}
\newcommand{\nhatsub}[1]{\hat{n}_{ #1}}
\newcommand{\Dsubsup}[2]{D_{\scriptscriptstyle #1}^{\scriptscriptstyle (#2)}}
\newcommand{\Tsubsup}[2]{T_{\scriptscriptstyle #1}^{\scriptscriptstyle (#2)}}
\newcommand{\vDsup}[1]{\mathbf{D}^{\scriptscriptstyle (\!#1\!)}}
\newcommand{\vTsup}[1]{\mathbf{T}^{\scriptscriptstyle (\!#1\!)}}
\newcommand{\vZsup}[1]{\mathbf{Z}^{\scriptscriptstyle (\!#1\!)}}
\newcommand{\vCsup}[1]{\mathbf{C}^{\scriptscriptstyle (\!#1\!)}}
\newcommand{\Omsub}[1]{\Omega_{ #1}}
\newcommand{\Lsub}[1]{L_{ #1}}
\newcommand{\tausub}[1]{\tau_{\scriptscriptstyle #1}}

\newcommand{\omsub}[1]{\omega_{ #1}}
\newcommand{\zetasub}[1]{\zeta_{\scriptscriptstyle  #1}}
\newcommand{\Phisub}[1]{\Phi_{\scriptscriptstyle \! #1}}
\newcommand{\phisub}[1]{\phi_{\scriptscriptstyle #1}}
\newcommand{\thetasub}[1]{\theta_{\scriptscriptstyle #1}}

\newcommand{\psisub}[1]{\psi_{\scriptscriptstyle #1}}

\newcommand{\Fevac}{F_{\!\text{ev}}}

\newcommand{\mevac}{m_{\text{ev}}}
\newcommand{\Nevac}{N_{\text{ev}}}
\newcommand{\PmcS}{\mathcal{P}}
\newcommand{\Psub}[1]{P_{\! #1}}
\newcommand{\gllth}{g_{\text{th}}^{\scriptscriptstyle(\! 2\!)}}
\newcommand{\gll}{g^{\scriptscriptstyle(\! 2\!)}}
\newcommand{\tsub}[1]{t_{#1}}
\newcommand{\eqlab}[1]{\label{eq:#1}}
\renewcommand{\eqref}[1]{Eq.~(\ref{eq:#1})}
\newcommand{\eqsref}[2]{Eqs.~(\ref{eq:#1}) and~(\ref{eq:#2})}

\newcommand{\secref}[1]{Section~\ref{sec:#1}}
\newcommand{\secsref}[2]{Sections~\ref{sec:#1} and~\ref{sec:#2}}

\newcommand{\figref}[1]{Fig.~\ref{fig:#1}}

\newcommand{\figlab}[1]{\label{fig:#1}}
\newcommand{\equal}{\!=\!}
\newcommand{\minus}{\!-\!}
\newcommand{\plus}{\!+\!}
\newcommand{\mytimes}{\!\times\!}

\newcommand{\kapS}{\kappa_{s}}
\newcommand{\kapL}{\kappa_{\scriptscriptstyle L}}
\newcommand{\DelI}{\Delta_{i}}
\newcommand{\DelS}{\Delta_{s}}
\newcommand{\om}{\omega}
\newcommand{\Sin}{S_{\text{pump}}}
\newcommand{\ssub}[1]{{s}_{ #1}}
\newcommand{\nsup}[1]{{n}^{\scriptscriptstyle (\!#1\!)}}
\newcommand{\nsubsup}[2]{n_{ #1}^{\scriptscriptstyle (\!#2\!)}}
\newcommand{\Ssup}[1]{{S}^{\scriptscriptstyle (\!#1\!)}}
\newcommand{\Isup}[1]{{I}^{\scriptscriptstyle (\!#1\!)}}
\newcommand{\xsup}[1]{{d}^{\scriptscriptstyle (\!#1\!)}}

\newcommand{\psup}[1]{{p}^{\scriptscriptstyle (\!#1\!)}}

\newcommand{\vxsup}[1]{{\boldsymbol{d}}^{\scriptscriptstyle (\!#1\!)}}
\newcommand{\vpsup}[1]{{\boldsymbol{p}}^{\scriptscriptstyle (\!#1\!)}}
\newcommand{\vxsubsup}[2]{{\boldsymbol{d}}_{ #1}^{\scriptscriptstyle (\!#2\!)}}
\newcommand{\mcPtot}{\mathcal{P}_{\text{tot}}}

\newcommand{\Fth}{F_{\text{th}}}

\newcommand{\ahat}{\hat{a}}
\newcommand{\mcX}{\mathcal{X}}
\newcommand{\taubin}{\tau_{\scriptscriptstyle \text{bin}}}
\newcommand{\tD}{\tau_{\scriptscriptstyle D}}
\newcommand{\tM}{t_{\scriptscriptstyle M}}
\newcommand{\Qintr}{Q_{\scriptscriptstyle L}}
\newcommand{\ket}[1]{|{#1}\rangle}

\newcommand{\expect}[1]{\langle {#1}\rangle}
\newcommand{\braket}[2]{ \left \langle #1 | #2 \right \rangle}
\newcommand{\ic}{\text{i}}
\newcommand{\nn}{\nonumber}
\newcommand{\Ntraj}{N_{\text{traj}}}

\begin{document}


\title{Temporally and spectrally multiplexed single photon source using quantum feedback control for scalable photonic quantum technologies\\}

\author{Mikkel Heuck$^{1,2,*}$, Mihir Pant$^{2}$, Dirk R. Englund$^{2,\dagger}$}
\address{${}^{1}$Department of Photonics Engineering, Technical University of Denmark, Building 343, 2800 Kgs. Lyngby, Denmark\\
${}^2$Department of Electrical Engineering and Computer Science, Massachusetts Institute of Technology,~77 Massachusetts Avenue, Cambridge, Massachusetts 02139, USA\\
}
\ead{${}^{*}$mheu@fotonik.dtu.dk, ${}^{\dagger}$englund@mit.edu}

\begin{abstract}
Current proposals for scalable photonic quantum technologies require on-demand sources of indistinguishable single photons with very high efficiency. Even with recent progress in the field there is still a significant gap between the requirements and state of the art performance. Here, we propose an on-chip source of time-multiplexed, heralded photons. Using quantum feedback control on a photon storage cavity with an optimized driving protocol, we estimate an on-demand efficiency of $99\%$ and unheralded loss of order $1\%$, assuming high efficiency detectors and intrinsic cavity quality factors of order $10^8$. We further explain how temporal- and spectral-multiplexing can be used in parallel to significantly reduce device requirements if single photon frequency conversion is possible with efficiency in the same range of $99\%$.
\end{abstract}


\section{\label{sec:introduction} Introduction}
Achieving sources of on-demand pure single photon states has been a long-standing goal of quantum information science~\cite{Kok2007}. Recent years have seen considerable progress in the performance of `deterministic sources' based on two-level quantum emitters~\cite{Somaschi2015, Ding2016, Aharonovich2016}. Additionally, the efficiency of sources based on probabilistic processes, such as parametric down-conversion and spontaneous four-wave mixing (sFWM), has been improved by multiplexing either spatial~\cite{Collins2013}, temporal~\cite{Kaneda2015a, Glebov2013}, or spectral~\cite{Joshi2017} degrees of freedom of photons.
Despite this progress, a large gap remains between state-of-the-art demonstrations and the requirements of proposed quantum information processing technologies, including photonic quantum repeaters~\cite{pant2017rate}, precision sensors~\cite{Giovannetti2011}, and photonic quantum computing~\cite{li2015, pant2017percolation}. 
We believe this calls for investigations into novel device concepts that are necessary to bridge this gap. 

In this work, we investigate the feasibility of single photon sources that meet the requirements of scalable photonic quantum technologies: near-unity purity single photons produced in a reproducable chip-integrated photonic circuit. Our proposal uses temporal multiplexing of parametrically produced signal-idler photon pairs and includes the possibility of additional multiplexing of the spectral degree of freedom leading to significantly improved performance. 

We consider a control protocol based on Bayesian inference with both idler and signal photon detection to optimize the signal-photon state. This approach shows the trade-off between heralding the generation of a single photon state and its purity. Our study reveals that, for near-term realistic device parameters, highly efficient ($\sim \!99$\%) sources of single photons could be possible in scalable nanophotonic platforms. As illustrated in ~\figref{struct}a, our proposed device  consists of a high-$Q$ microring resonator ($Q$ of 10-100 million) consisting of a material, such as silicon, with a $\chi^{(3)}$ nonlinearity for photon-pair generation by sFWM. This storage ring is coupled to photon number resolving detectors (PNRDs) through Mach-Zehnder interferometer (MZI) filters~\cite{Barbarossa1995}. The filters enable decoupling of certain frequencies from the waveguide by controlling the path-imbalance of the MZI relative to the length of the ring (see~\figref{struct}b). Idler photons and the pump field couple out of the storage ring within a single time bin, whereas the signal can be stored for up to $M$ bins.
\begin{figure}[!htb]
  \centering
  \includegraphics[height=4cm]{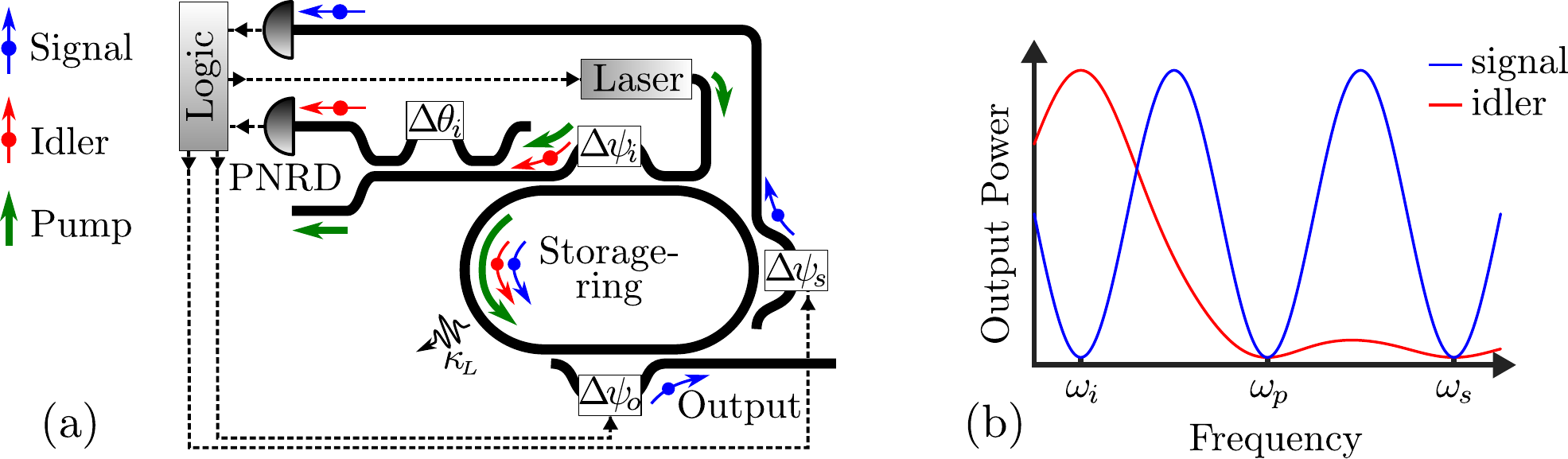}
 \caption{(a) Storage- and release design. Solid lines are optical waveguides, while dashed lines represent electrical control signals. PNRD: photon number resolving detector. (b) Illustration of the power spectrum coupled out of the signal filter in its closed configuration and the spectrum arriving at the idler detector (see~\secref{device architecture}). 
}
\figlab{struct}
\end{figure}
The signal- and output filters contain tunable phases, $\Delta\psi_s(t)$ and $\Delta\psi_o(t)$, allowing them to dynamically couple out signal photons to a detector or output waveguide, respectively. Each emission cycle is divided into $M$ time bins in which we either: 1) Pump to generate a photon pair; 2) release excess signal photons by tuning the phase of the signal filter; 3) evacuate all photons from the system through the signal filter; or 4) store the signal state if detection events suggest a single photon is present. The driving protocol prescribes which action is taken in a given time bin depending on the information available from detection events. We optimize the protocol to maximize the probability of a single signal photon occupying the storage ring at the emission time, $\tsub{M}$. Signal photons are emitted by tuning the output filter to its open state. Tailoring the temporal shape of $\Delta\psi_o(t)$ allows shaping the output photon wavepacket. Note that decoupling the signal mode from the environment reduces the spectral correlations of the signal-idler quantum state, which increases the purity of the single photon state of the signal after detecting the idler.\\


This article is organized as follows: \secref{device architecture} details the device architecture and explains how multiplexing in both time- and frequency is possible.~\secsref{model}{probability analysis} present the model and probability analysis for evaluating the performance of our proposed architecture and~\secref{protocol} discusses the  driving protocol. \secref{results} presents simulation results and~\secref{discussion} concludes with a discussion of the feasibility of experimental demonstrations.

\section{\label{sec:device architecture} Device architecture}
Our proposed device implementation uses a photonic integrated circuit consisting of a ring resonator and MZI-couplers~\cite{Barbarossa1995}, as illustrated in~\figref{struct}a. The structure is shown again in~\figref{gate ring struct} with definitions of fields used in the analysis. 
\begin{figure}[!htb]
  \centering
  \includegraphics[height=4.5cm]{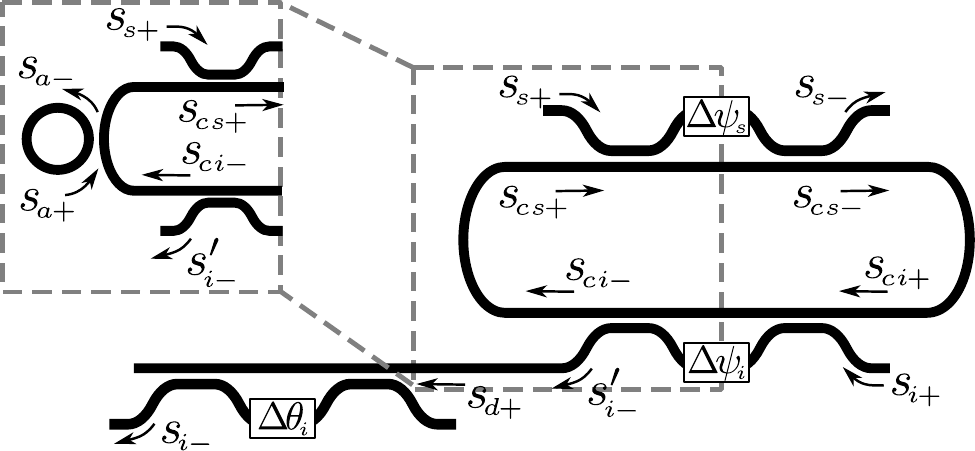}
 \caption{Device architecture including definitions of fields used in the analysis. Note that the output filter is omitted. The inset shows the addition of an auxiliary ring to enable high efficiency frequency conversion (see~\secref{frequency conversion}).}
\figlab{gate ring struct}
\end{figure}
The outputs of the MZI filters are related to the inputs by
\begin{align} \eqlab{mzi relation matrix}
\left[\begin{array}{c}\ssub{n-}		\\ 
\ssub{cn-} \end{array} \right] &= 
\vCsup{n} \vZsup{n} \vCsup{n}
\left[\begin{array}{c} \ssub{n+}		\\ 
\ssub{cn+} \end{array} \right]\!,~ n=\{i,s\}
\end{align}
where $n\equal \{i,s\}$ represent the idler and signal filter. The matrices $\vCsup{n}$ and $\vZsup{n}$ are given by
\begin{align}\eqlab{symmetric beam splitter}
\hspace{-1mm} \vCsup{n}  \equal 
\!\left[\!\begin{array}{c c} \nu_n  	& \ic\sqrt{1-\nu_n^2} \\
\ic\sqrt{1-\nu_n^2} 	& \nu_n  \end{array} \!\right] \!\!, ~
\vZsup{n}  \equal 
\!\left[\!\begin{array}{c c} e^{\ic\psisub{nT}} 	& 0 \\
0 	& e^{\ic\psisub{nB}}  \end{array} \!\!\right] \!\!,
\end{align}
where $\psisub{nT}$ and $\psisub{nB}$ are the phase accumulation in the arm containing the phase shifter and the arm that is part of the ring, respectively. The through-coupling of the waveguide couplers is $\nu_n$. The transfer matrix $\vTsup{n}\equal \vCsup{n}\vZsup{n}\vCsup{n}$ of the MZI filter is 
\begin{align} \eqlab{transfer matrix}
\hspace{-1.5mm} \vTsup{n} \equal  e^{\ic\psisub{nB}} \!\!
\left[ \!\! \begin{array}{c c} (1\plus e^{\ic\psisub{n}}\!)\nu_n^2\minus 1 	&  \!\! \ic(1\plus e^{\ic\psisub{n}}\!)\nu_n\sqrt{1\minus\nu_n^2} \\ 
\ic(1\plus e^{\ic\psisub{n}}\!)\nu_n\sqrt{1\minus\nu_n^2} 	& \!\! \nu_n^2 \minus e^{\ic\psisub{n}}(1\minus\nu_n^2)	 \end{array} \!\! \right]  \!\!,
\end{align}
where $\psisub{n}\equal \psisub{nT} - \psisub{nB}$ is the difference in phase accumulation between the two arms. We assume that the phases $\Delta\psisub{n}$ are tunable such that
\begin{align}\eqlab{psi n}
\psisub{n}(\om) &= k(\om) \Delta \Lsub{n}  + \Delta\psisub{n} , ~ n=\{i,s\}.
\end{align}
Here, the path length difference between the MZI arms is $\Delta\Lsub{n}$ and the propagation constant is approximated as 
\begin{align}
k(\om) \approx  \frac{\nefft}{c}\om_0 +  \frac{n_g}{c} (\om-\om_0),
\end{align}
where the complex effective mode index is $\nefft \equal \neff' \plus \ic\neff''$ and the group index is defined from $n_g/c\equiv \partial k/\partial  \om $. \\

As mentioned in~\secref{introduction} the filters must be designed to only allow certain frequencies to pass. To illustrate how this may be accomplished let us consider a situation where a field, $s_f$, is generated inside the ring between the signal and idler filter, such that 
\begin{align} \eqlab{internally fed ring}
\ssub{ci+} = e^{\ic \phi_{si}} \ssub{cs-} + \ssub{f} , \qquad \ssub{s+} = \ssub{i+} = 0.
\end{align}
The fields are related using~\figref{gate ring struct} and~\eqref{transfer matrix}
\begin{align} \eqlab{sci+}
\ssub{ci+} = e^{\ic \phi_{si}}\Tsubsup{2,2}{s} \ssub{cs+} + \ssub{f} =
e^{\ic \phi_{si}}\Tsubsup{2,2}{s} e^{\ic \phi_{is}}\ssub{ci-} + \ssub{f} =
e^{\ic \phi_{si}}\Tsubsup{2,2}{s} e^{\ic \phi_{is}}\Tsubsup{2,2}{i} \ssub{ci+} + \ssub{f}, 
\end{align}
where $\Tsubsup{i,j}{n}$ is the matrix element of $\vTsup{n}$ corresponding to the $i$th row and $j$th column. The round-trip phase of the isolated storage ring is
\begin{align}\eqlab{round trip phase isolated ring}
\phisub{c}(\om) = \psisub{iB} + \phi_{is} + \psisub{sB}+ \phi_{si}  = k(\om)\Lsub{c},
\end{align}
where $\Lsub{c}$ is the length of the storage ring. From~\eqref{sci+}, we have  
\begin{align} \eqlab{sci+ sol}
\ssub{ci+} &= \frac{1}{1 - e^{\ic\phisub{c}}\zetasub{i}\zetasub{s} } \ssub{f},
\end{align}
where $\zetasub{n}$ is a tuning parameter of the ring-waveguide coupling given by
\begin{align} \eqlab{zeta}
\zetasub{n} \equiv \Tsubsup{2,2}{n} e^{-\ic\psisub{nB}} =  \nu_n^2 \minus e^{\ic\psisub{n}}(1-\nu_n^2).
\end{align}
%
The out-going fields are given by
%
\begin{align} \eqlab{s out vs T}
\ssub{s-} = \Tsubsup{1,2}{s} \ssub{cs+} =  \frac{ \Tsubsup{1,2}{s}e^{\ic\phisub{is}} \Tsubsup{2,2}{i} }{1 - e^{\ic\phisub{c}}\zetasub{i}\zetasub{s} } \ssub{f} , \qquad 
\ssub{i-} = \Dsubsup{1,2}{i} \ssub{i-}'  =  \frac{ \Dsubsup{1,2}{i} \Tsubsup{1,2}{i} }{1 - e^{\ic\phisub{c}}\zetasub{i}\zetasub{s} } \ssub{f} .
\end{align}
%
The drop filter transfer matrix is
\begin{align}\eqlab{D matrix}
\vDsup{\;i\;} =  e^{\ic\thetasub{iB}}  \frac12 \!\!
\left[ \!\! \begin{array}{c c}  e^{\ic\thetasub{i}} \minus 1 	&   \ic(1\plus e^{\ic\thetasub{i}}) \\ 
\ic(1\plus e^{\ic\thetasub{i}})	& 1\minus e^{\ic\thetasub{i}}	 \end{array} \!\! \right] .
\end{align}
The through-coupling coefficient, $\nu_n$, from \eqref{symmetric beam splitter} was chosen to be $1/\sqrt{2}$ in~\eqref{D matrix} to achieve 100\% visibility of the drop filter. The phase difference between the arms is
\begin{align}\eqlab{Psi n} 
\thetasub{i}(\om) = k(\om)\Delta\Lsub{di} + \Delta\thetasub{i} .
\end{align}
The idler, pump, and signal frequencies are chosen from three adjacent modes of the storage ring, such that
\begin{align} \eqlab{mode freq props}
\omsub{i} = \omsub{p} - \Omsub{c}, \qquad 
\omsub{s} = \omsub{p} + \Omsub{c},
\end{align}
where $\Omsub{c}$ is the free spectral range (FSR) of the storage ring.
From the out-coupling matrix elements
\begin{align} \eqlab{T1,2}
\Tsubsup{1,2}{n} = \Tsubsup{2,1}{n} = e^{\ic\psisub{nB}} \ic(1\plus e^{\ic\psisub{n}})\nu_n\sqrt{1\minus\nu_n^2},
\end{align}
it is observed that frequencies corresponding to $\psisub{n}(\om) \equal \pi + 2\pi p$ cannot pass through the filters. This fact is used to realize a way to obtain the desired properties of the filters by chosing:
\begin{subequations}\eqlab{filter props}
\begin{align} 
\psisub{i}(\omsub{i}) &= 2\pi, ~& \psisub{s}(\omsub{i}) &= \pi  , ~& \thetasub{i}(\omsub{i}) &= \pi,\\
\psisub{s}(\omsub{s}) &= \pi, ~&  \psisub{i}(\omsub{s}) &= \pi	, ~& \thetasub{i}(\omsub{p}) &= \pi.
\end{align}
\end{subequations}
%
The conditions in~\eqref{filter props} can be met by making the FSRs of the filters different integer values of $\Omsub{c}$. Correspondingly, the path length differences should be different integer fractions of $\Lsub{c}$ as 
\begin{align}\eqlab{L props}
\Delta\Lsub{i} = \Lsub{c}/4, ~~ \Delta\Lsub{di} = \Lsub{c}/2,  ~~ \Delta\Lsub{s}  = \Lsub{c} .
\end{align}
In~\figref{gate prop 1}a we plot the output fields (top panel) and the field circulating inside the storage ring (bottom panel) for these choices. From the top panel it is seen that $|\ssub{i-}|^2$ vanishes at $\omsub{p}$ and $\omsub{s}$, and $|\ssub{s-}|^2$ goes to zero at $\omsub{i},~\omsub{p}$, and $\omsub{s}$. We also plot the field in the drop port, $\ssub{d-}\equal  \Dsubsup{1,1}{i}\ssub{i-}'$, to show that it has no contributions from the signal and idler frequencies.
\begin{figure}[!htb] 
  \centering
  \includegraphics[height=4.5cm]{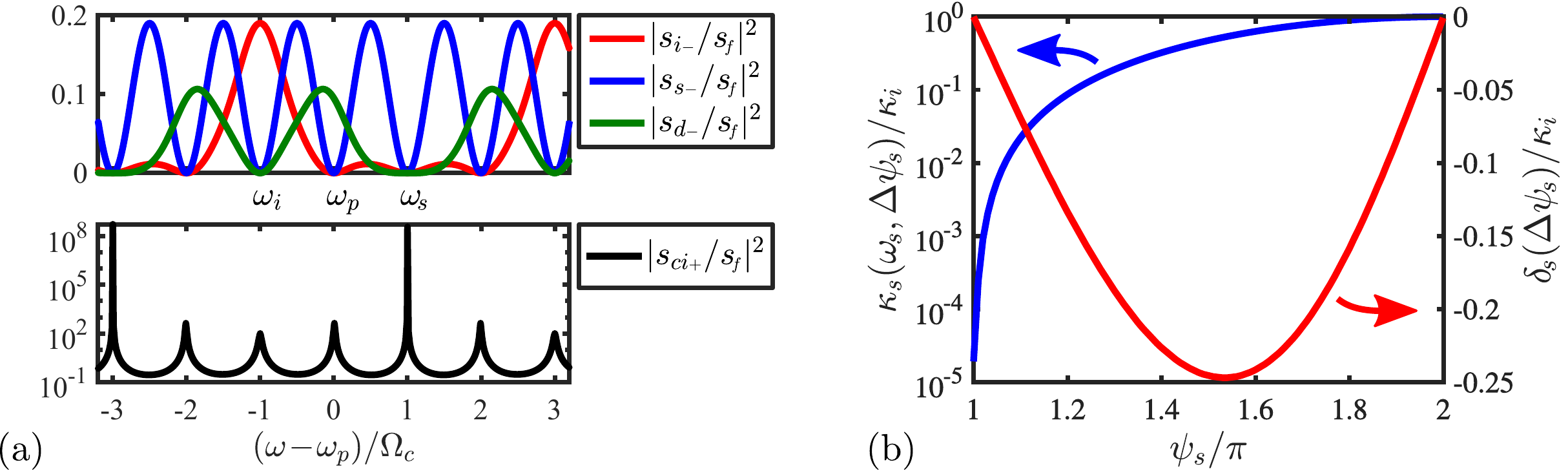}
 \caption{(a) Top: Output fields, $|\ssub{n-}\big / \ssub{\!f}|^2$, as a function of frequency. Bottom: Circulating field $|\ssub{ci+} \big / \ssub{\!f}|^2$ as a function of frequency. (b) Coupling rate and resonance shift as a function of $\psisub{s}$. Both curves are normalized by the static coupling rate through the idler filter, $\kapsub{i}(\omsub{i})$. Parameters: $\nu^2\equal 0.95$, $\Lsub{c}\equal 100\mu$m, $\neff''\equal 10^{-7}$, $\neff'\equal 2.5$, and $n_g\equal 4$. } 
\figlab{gate prop 1} 
\end{figure}
From the bottom panel of~\figref{gate prop 1}a, it is observed that the signal mode at $\omsub{s}$ is spectrally narrow compared to the idler and pump modes. This is caused by the choice $\psisub{s}(\omsub{s})\equal \pi$, which corresponds to the signal filter being closed and the $Q$-factor only being limited by intrinsic loss. The design choices in~\eqref{L props} are thus seen to yield the desired filter properties.

The signal filter is tuned by modifying $\Delta\psisub{s}$. The corresponding change in the cavity-waveguide coupling is found from the tuning parameter $\zetasub{s}$ in~\eqref{zeta}. The amplitude of the matrix element $\Tsubsup{2,2}{n}$ describes the loss per round trip of the intra-cavity field due to waveguide coupling. From~\eqref{zeta} it is therefore seen that a coupling rate may be defined by $\exp\!\!\big[\!-\!\kapsub{n} \tausub{\rm{RT}}\big]\equal |\zetasub{n}|$, or
\begin{align} \eqlab{gamW}
\kapsub{n}(\om,\Delta\psisub{n} ) = -\frac{c}{n_g\Lsub{c}}\ln\!\big[ |\zetasub{n}(\om,\Delta\psisub{n}) | \big],
\end{align}
where $ \tausub{\rm{RT}}$ is the cavity round-trip time. 
The resonances, $\om_n$, of the MZI-coupled ring are also affected by tuning $\Delta\psisub{s}$. They are found by the resonance condition on the round-trip phase of the MZI-coupled ring
\begin{align}\eqlab{resonance condition}
\Phisub{n}(\om_n)  =\phisub{c}(\om_n) + \arg\!\big[\zetasub{i}(\om_n)\zetasub{s}(\om_n) \big] =  2\pi .
\end{align}
In~\figref{gate prop 1}b we plot the coupling rate, $\kapsub{s}$, and resonance shift
\begin{align}\eqlab{resonance shift}
\delta_s(\psisub{s})\equal \omsub{s}(\psisub{s}) - \omsub{s}(\psisub{s}\equal \pi)
\end{align}
as a function of $\psisub{s}$. The coupling rate remains non-zero at $\psisub{s}\equal \pi$ due to the small imaginary part of the complex refractive index.\\

If the tunable phase, $\Delta\psisub{s}(t)$, has some time variation due to the electrical signal coming from the logic unit (see~\figref{struct}a), the coupling rate will also vary in time, $\kapsub{s}(t)$. The interference-based filtering of the MZI will work as long as $\Delta\psisub{s}(t)$ varies slowly relative to the propagation time through the MZI. In~\secref{model} we use this fact to model the system as a single resonator with the signal mode having a time-dependent coupling rate. The resonance shift, $\delta_s$, will be neglected by assuming that the chirp it induces on the signal photon does not affect its detection.


\subsection{\label{sec:frequency conversion} Frequency conversion}
Since the storage ring has many modes separated by a FSR, signal and idler pairs are generated in spectral modes symmetrically distributed about the degenerate pump mode. By heralding on multiple idler modes, multiplexing in the frequency domain in addition to the time domain is possible~\cite{Joshi2017}. With the choices of $\Delta\Lsub{n}$ made here, any mode-pair satisfying  
\begin{align}\eqlab{all usable modes}
\omsub{s} & \equal \omsub{p} \plus (1\plus 4p)\Omsub{c}, \quad \omsub{i} \equal \omsub{p} \minus (1\plus 4p)\Omsub{c}, ~p\in \mathbb{Z}_0
\end{align}
could be used. Multiplexing in frequency requires frequency-conversion of signal photons to the target wavelength~\cite{Joshi2017, Li2016}. If Bragg-Scattering FWM~\cite{McKinstrie2005} is used for the frequency-conversion, the probability of up-conversion and down-conversion is equal (provided that phase-matching is uniform across several FSRs of the storage ring)~\cite{Li2016, Vernon2016}. To overcome this symmetry, the storage ring can be coupled to an auxiliary ring with a length $L_r\equal L_c/16$ such that its resonances coincide with every fourth signal mode (see~\figref{gate ring struct}b). The ring-ring coupling will cause a splitting of the mode spectrum and thereby effectively eliminate either the up- or down-conversion cavity mode. To illustrate this, let us consider the auxiliary ring being coupled on the left side of the storage ring with a coupling region described by a matrix $\vCsup{a}$ as in~\eqref{symmetric beam splitter} (see~\figref{gate ring struct}b). The fields are then related by
%
\begin{align} \eqlab{aux ring coupling}
\ssub{cs+} = \nu_a \ssub{ci-} + \ic\sqrt{1-\nu_a^2} \ssub{a+}, \quad
\ssub{a-} = \ic\sqrt{1-\nu_a^2} \ssub{ci-} + \nu_a \ssub{a+}, \quad
\ssub{a+} = e^{\ic\phisub{a}}\ssub{a-},
\end{align}
%
where $\nu_a$ is the coupling coefficient of the directional coupler formed by the waveguide sections of each ring, and $\phisub{a}$ is the round-trip phase of the auxiliary ring. Solving~\eqref{aux ring coupling} yields
\begin{align} \eqlab{aux ring trans}
\ssub{cs+} &=  \left(\nu_a - \frac{(1-\nu_a^2)e^{\ic\phisub{a}}}{1-\nu_a e^{\ic\phisub{a}}}  \right)\ssub{ci-} .
\end{align}
Inserting~\eqref{aux ring trans} into~\eqref{sci+} leads to a modified version of~\eqref{zeta}
\begin{align} \eqlab{sci+ mod} 
\ssub{ci+} &= \frac{1}{1 - e^{\ic\phisub{c}}\zetasub{i}\zetasub{s} \left(\nu_a - \frac{(1-\nu_a^2)e^{\ic\phisub{a}}}{1-\nu_a e^{\ic\phisub{a}}}  \right)} \ssub{f}.
\end{align}
In~\figref{freq multiplex} we show the circulating field with- and without the auxiliary ring coupled. 
\begin{figure}[!htb] 
  \centering
  \includegraphics[height=4.5cm]{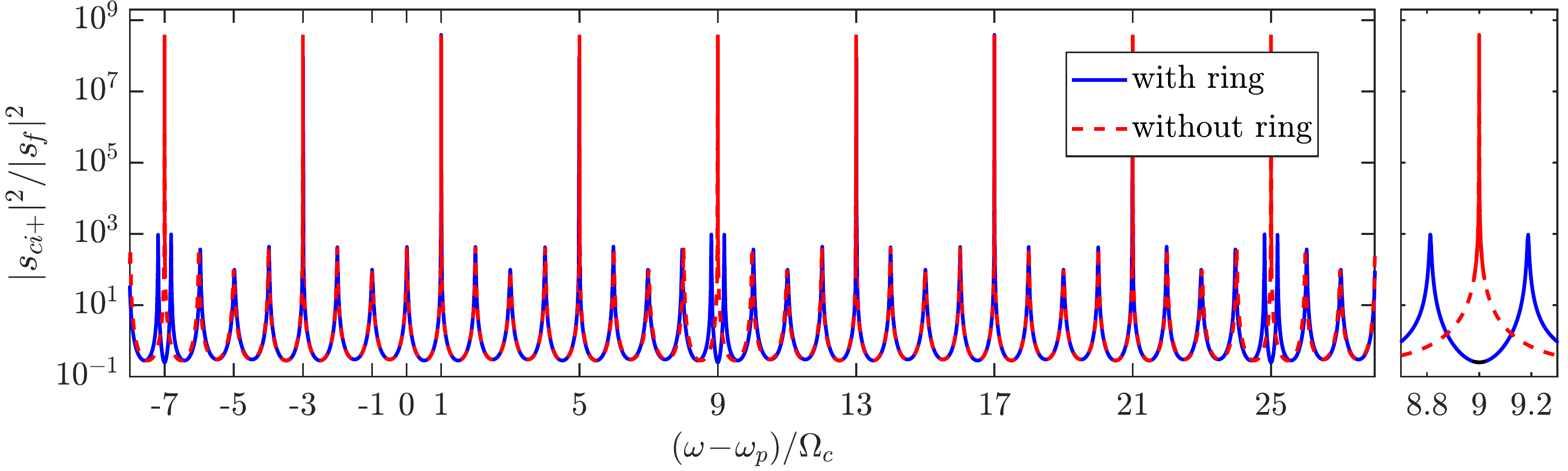}
 \caption{Power spectrum of the circulating field in the storage ring with- (solid blue) and without (dashed red) an auxiliary coupled ring. Parameters are the same as in~\figref{gate prop 1}, and $\nu_a^2\equal 0.9$. } 
\figlab{freq multiplex} 
\end{figure}
Notice the splitting of the cavity modes at the suppressed frequencies
\begin{align} \eqlab{blocked freqs}
\omega_{\rm{supp}} =  \omsub{p} \plus (9\plus 16p)\Omsub{c}, ~p\in\mathbb{Z}_0 .
\end{align}
Both the storage- and auxiliary ring have a resonance at these frequencies and their coupling gives rise to two super-modes that are shifted away from the original resonances.

Detecting an idler photon at e.g. $\omsub{i}\equal \omsub{p}-5\Omsub{c}$ heralds the presence of a signal photon at $\omsub{s}\equal \omsub{p}+5\Omsub{c}$. Since the storage ring mode at $\omsub{p}+9\Omsub{c}$ is now shifted, the signal can be down-converted to the target frequency $\omsub{p}+\Omsub{c}$ with high efficiency. In general, we can frequency-multiplex using all signal modes satisfying the relation
\begin{align} \eqlab{freq shift modes}
\omsub{s} =  \omsub{p} \plus (5\plus 8p)\Omsub{c}, ~p\in\mathbb{Z}_0.
\end{align}
The fact that the signal photon is born inside the storage cavity (as opposed to the situation in Refs.~\cite{Joshi2017, Li2016}) significantly simplifies the problem and near-unity conversion efficiency should be possible~\cite{Huang2013, Vernon2016}.

If $\PmcS$ is the success probability of each frequency mode the total success probability from using $N_{\!F}$ modes is $\mcPtot \equal 1 \minus  (1\minus \PmcS)^{N_{\!\scriptscriptstyle F}}$, assuming perfect conversion efficiency and no reduction in temporal multiplexing efficiency of each frequency channel. 

\section{\label{sec:model} Temporal multiplexing model}
The number of time bins available for multiplexing depends on the intrinsic decay rate of the storage ring, $\kapL$, and the speed of the feedback controls. The $m$th time bin is defined by the time interval $[\tsub{m\minus 1},~\tsub{m}]$ and all bins are assumed to be of equal length, $\taubin\equal \tsub{m}\minus \tsub{m\minus 1}$.
The feedback controls are the pump power, $|\Sin(t)|^2$, and signal filter phase, $\Delta\psisub{s}(t)$. 
The processing time of the logic unit, $\tD$, determines the necessary lag between the time of deciding the action in bin $m\plus 1$, $\tsub{m^{\!*}}$, and its onset, $\tsub{m} \equal \tsub{m^{\!*}} + \tD$. If $N_I(\tsub{m})$ and $N_S(\tsub{m})$ denote the number of idler and signal detections up until $\tsub{m}$, the detection number is defined as
\begin{align} \eqlab{x def}
\xsup{m} \equiv  N_I(\tsub{m}) - N_S(\tsub{m}) .
\end{align}
We infer the state of the storage ring at $\tsub{m}$ based on the value of the detection number at the decision time, $\tsub{m^{\!*}}$. For instance, $\xsup{m^{\!*}}\equal 1$ suggests that one signal photon occupies the cavity at $\tsub{m}$, which we denote as $\nsup{m}\equal 1$. 
If $\xsup{m^{\!*}} \! \geq\! 0$, the estimated number of signal photons is $\nsubsup{\scriptscriptstyle\rm{est}}{m} \equal \xsup{m^{\!*}}$. Since some idler photons might not be detected, the detection number can be negative and in this case we always pump the cavity and our estimation is therefore $\nsubsup{\scriptscriptstyle\rm{est}}{m}\equal \xsup{m^{\!*}} \minus \xsup{m\minus 1}$.

The state estimation fidelity is the probability that our estimate of the state is correct and is given by
\begin{align} \eqlab{state fidelity}
P(\nsup{n}\equal \nsubsup{\scriptscriptstyle\rm{est}}{m} | \vxsup{m^{\!*}} ) =  \frac { P(\nsup{n} \equal \nsubsup{\scriptscriptstyle\rm{est}}{m}, \vxsup{m^{\!*}} ) } {P( \vxsup{m^{\!*}} ) }.
\end{align}
The detection sequence $\vxsup{m^{\!*}}\equiv \{ \xsup{0},\xsup{1},\ldots, \xsup{m\minus1}, \xsup{m^{\!*}} \}$ contains information from all previous bins. We consider the photon generation successful only if the heralding efficiency (probability of having a single photon in the cavity conditioned on the given detection sequence) and the second-order correlation obey the threshold conditions
\begin{align} \eqlab{criteria}
 P(\nsup{M} \equal 1  | \vxsup{M^{\!*}}) \geq \Fth ~~\text{and} ~~ 
\gll(\vxsup{M^{\!*}}) \leq \gllth,
\end{align}
where $M$ enumerates the last bin of the emission cycle and $\gll(\vxsup{M^{\!*}})$ is defined by~\cite{Li2016a}
\begin{align} \eqlab{g2}
\gll(\vxsup{M^{\!*}}) \equal \frac{ \displaystyle \sum_{\nsup{M}} \nsup{M}(\nsup{M}\minus1)P(\nsup{M} | \vxsup{M^{\!*}}) }{\bigg[ \displaystyle \sum_{\nsup{M}} \nsup{M} P(\nsup{M} | \vxsup{M^{\!*}}) \bigg]^2} .
\end{align}
The thresholds $\Fth$ and $\gllth$ are performance metrics of the source and can be chosen according to any application of interest. The success probability (probability that exactly one signal photon occupies the storage ring at $\tsub{M}$) is
\begin{align}\eqlab{Psucc}
\PmcS(M) 	= \!\!\sum_{\vxsup{M^{\!*}}} \!P(\nsup{M}\equal 1, \vxsup{M^{\!*}}),
\end{align}
where the summation runs over all detection sequences fulfilling~\eqref{criteria}. \figref{bins} illustrates two examples of detection sequences leading to successful state preparation. In~\figref{bins}a, the detection of one idler photon in bin 4 heralds the presence of one signal photon and the state is stored until the end of the emission cycle.
\begin{figure}[!h]
  \centering
  \includegraphics[height=3.3cm]{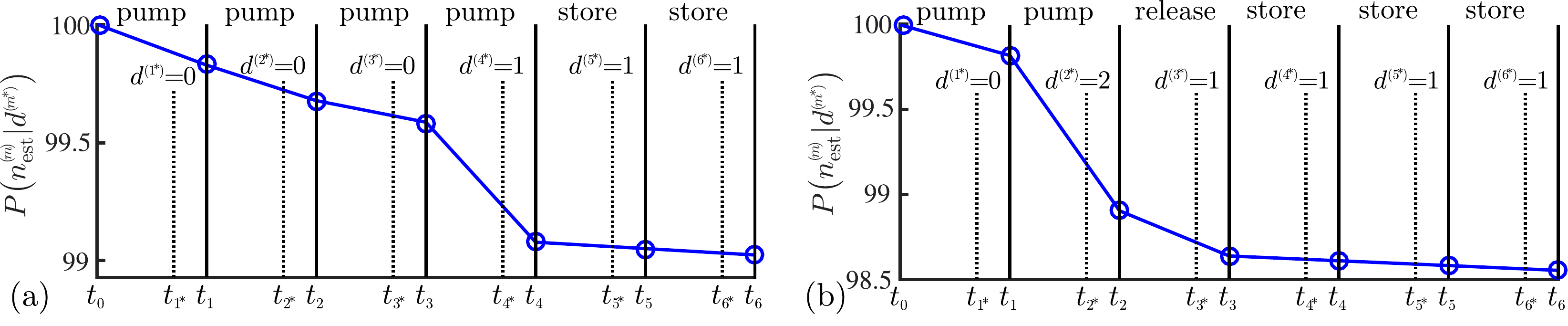}
 \caption{State estimation fidelity as a function of time for two detection sequences and $M\equal 6$. The actions prescribed by the driving protocol (see~\secref{protocol}) are listed for each bin. Parameters are the same as in~\figref{results 1}.} 
\figlab{bins}
\end{figure}
In~\figref{bins}b, the detection of two idler photons in bin 2 leads to release of signal photons in bin 3. The detection of one signal photon in bin 3 suggests that the desired state is achieved and it is stored. \\

The system in~\figref{struct}a can be modeled by considering three modes of the storage ring where the control-phase of the signal filter, $\Delta\psisub{s}(t)$, is represented by a time-dependent coupling rate, $\kapS(t)$, for the signal mode as discussed in~\secref{device architecture}. Photon pair generation is modeled using a Hamiltonian of the form
\begin{align}\eqlab{Hsys main}
H_{sys} 	=  \DelI \ahat_{i}^\dagger \ahat_{i} + \DelS \ahat_{s}^\dagger \ahat_{s}  + 
 \mcX \big( \ahat_{i}^\dagger \ahat_{s}^\dagger + \ahat_{i}\ahat_{s} \big),
\end{align}
where $\ahat_{s}$ and $\ahat_{i}$ are annihilation operators of the signal and idler modes, respectively. We use normalized units ($\hbar\equal 1$) and a classical pump rate described by $\mcX$, which is proportional to the nonlinear coefficient, $\chi^{\scriptscriptstyle (3)}$, and the energy of the pump cavity mode. Additionally, $\DelS$ and $\DelI$ are detunings between the pump frequency and the signal and idler modes, respectively. Coupling between the resonator and waveguides through the filters is modeled via collapse operators, $\Chatsub{nL} \equal \sqrt{2\kapL}\ahat_{n}$, and $\Chatsub{n} \equal\sqrt{2\kapsub{n}}\ahat_{n}$ with $n\equal i,s$~\cite{Johansson2013}. The loss rate, $\kapL$, of all modes is assumed equal. By neglecting self-induced nonlinear effects, the energy in the cavity follows from coupled mode theory~\cite{Haus1991}
\begin{align}\eqlab{X vs Pow}
\mcX(t) =  \left|\int_{-\infty}^{t} \!\! e^{-\kapsub{p}  (t-t')} \Sin(t') dt' \right|^2,
\end{align}
where the input power is assumed Gaussian, $|\Sin(t)|^2 \propto \exp[ -(t-\tsub{p})^2/\tau_p^2]$. The pump width $\tau_p$ is 1ps and the time $\tsub{p}$ is adjusted such that $\mcX(t_{m\minus 1})$ is at least a thousand times smaller than its peak value.  

The state of the storage ring, $\ket{\psi(t)}$, is calculated from~\eqref{Hsys main} using a Monte Carlo method~\cite{Johansson2013} with an initial state $\ket{\nsup{m\minus 1}, \nsubsup{ci}{m\minus 1} \equal 0}$. The assumption of zero idler photons at $\tsub{m\minus 1}$  ($\nsubsup{ci}{m\minus 1} \equal 0$) is based on the coupling rate, $\kapsub{i}$, being much larger than the inverse bin duration, $1/\taubin$. 

\section{\label{sec:probability analysis} Probability analysis}
We use Monte Carlo simulations~\cite{Johansson2013} to evaluate the probability distribution $P(\nsup{m}, \vxsup{m^{\!*}} )$. The assumptions are: 1) Detector dark counts are negligible. 2) The idler mode of the cavity is in the vacuum state at the beginning of each bin. 3) If $\xsup{m^{\!*}}\!\leq\! 0$, then $\nsup{m}\!\leq\! 2$, which is a good approximation for the large detection efficiency used in our simulations. 4) The signal coupling rate, $\kapS(t)$, can be varied without increasing the loss rate, $\kapL$, which is equal for all three modes. 

The probability $P(\nsup{m}, \vxsup{m^{\!*}})$ is evaluated using an expansion
\begin{align} \eqlab{herald m}
 P(\nsup{m}, \vxsup{m^{\!*}} ) =  \!\!\sum_{\nsup{m\minus 1}} \!\! P(\nsup{m}, \nsup{m\minus 1}, \vxsup{m^{\!*}} )  = 
 \!\!\sum_{\nsup{m\minus 1}} \!\! P(\nsup{m}, \xsup{m^{\!*}} | \nsup{m\minus 1}, \vxsup{m\minus 1} )  P(\nsup{m\minus 1}, \vxsup{m\minus 1} ) .
\end{align}
Notice that the probability distribution in the current bin $m$ is updated using information from the total duration of the previous bins, $\vxsup{m\minus 1}$, because the detector keeps acquiring information until the end of each bin. The second factor on the right-hand side (RHS) of~\eqref{herald m} is found by a similar expansion, 
which means that the distribution $P(\nsup{m}, \vxsup{m^{\!*}} )$ can be found iteratively starting from the first bin
\begin{align} \eqlab{herald 1}
P(\nsup{1}, \vxsup{1} ) = \sum_{\nsup{0}} P(\nsup{1}, \xsup{1} | \nsup{0}, \xsup{0} )  P(\nsup{0}, \xsup{0} ),
\end{align}
where $P(\nsup{0}, \xsup{0} )$ is known since $\nsup{0}$ and $\xsup{0}$ both equal zero at the beginning of each emission cycle. Note that we omit an expansion over the initial state of the idler mode in~\eqref{herald m} by assuming that it is in the vacuum state. 
In the following sections, it is explained how the probability distributions $ P(\nsup{m}, \xsup{m} | \nsup{m\minus 1}, \vxsup{m\minus 1} )$ and $ P(\nsup{m}, \xsup{m^{\!*}} | \nsup{m\minus 1}, \vxsup{m\minus 1} )$ are calculated for pumping and releasing, respectively.

\subsection{\label{app:pump model} Pumping}
If the cavity is pumped in bin $m$, the probability that there are $\nsup{m}$ signal photons in the cavity \emph{and} the detection number is $\xsup{m}$ at $t_m$ is
\begin{align} \eqlab{nm and xm given nm1 and xm1}
 \Psub{p}(\nsup{m}, \xsup{m} | \nsup{m\minus 1}, \vxsup{m\minus 1} ) = \!\!\!\! \sum_{\nsubsup{i}{m} = \Isup{m}}^{\infty}  \!\!\!\!  P(\xsup{m} |  \nsubsup{i}{m}) 
 \Psub{p}( \nsup{m},\nsubsup{i}{m} |\nsup{m\minus 1}, \vxsup{m\minus 1}),
\end{align}
where $\nsubsup{i}{m}$ is the number of idler photons coupled through the idler filter between $\tsub{m\minus 1}$ and $\tsub{m}$. The subscript $p$ is used to signify that we pump in bin $m$. The number of detected idler photons is $\Isup{m} \equal \xsup{m}- \xsup{m\minus 1}$, because the signal filter is closed so no signal detections contribute to $\xsup{m}$. Note that the probability that the detection number equals $\xsup{m}$ only depends on $\nsubsup{i}{m}$ and is given by
\begin{align} \eqlab{xm given nm1 vxm1}
P(\xsup{m} | \nsubsup{i}{m}) =  \left( \!\!\begin{array}{c}
\nsubsup{i}{m}\\
\Isup{m}
\end{array} \!\!\! \right) \eta^{\Isup{m}} (1\minus \eta)^{\nsubsup{i}{m}\minus \Isup{m}} ,
\end{align}
where $\eta$ is the detection efficiency. Since we only consider near-unity detection efficiency, we assume that the probability $P_p(\nsup{m\minus 1}\!\!>\!2 | \xsup{m^{\!*}\minus 1}\!\leq \!0 )$ is negligible and therefore truncate the summation in~\eqref{herald m} after 2 for bins where we pump (this is the third assumption in the beginning of this section). The distribution $ \Psub{p}(\nsup{m}, \xsup{m^{\!*}} | \nsup{m\minus 1}, \vxsup{m\minus 1} )$ is found by replacing $\xsup{m},~\nsubsup{i}{m}$, and $\Isup{m}$ by $\xsup{m^{\!*}},~\nsubsup{i}{m^{\!*}}$, and $\Isup{m^{\!*}}$ in~\eqsref{nm and xm given nm1 and xm1}{xm given nm1 vxm1}.


The probability of a certain configuration with $\nsup{m}$ photons in the signal mode and $\nsubsup{i}{m}$ idler photons in the detector waveguide is found by projecting the state $\ket{\nsup{m},  \nsubsup{ci}{m}}$ onto $\ket{\psi(t_m)}$ and tracing out the idler subspace for all Monte Carlo trajectories, where $\nsubsup{i}{m}$ idler photons couple into the detector waveguide
\begin{align}\eqlab{PsCiD}
\Psub{p}(\nsup{m}, \nsubsup{i}{m} | \nsup{m\minus 1}, \vxsup{m\minus 1} ) = \frac{1}{\Ntraj}\!\!
\sum_{\text{traj}(\nsubsup{i}{m})}  \sum_{\nsubsup{ci}{m} }    \left| \braket{\nsup{m},  \nsubsup{ci}{m} } {\psi(t_m)} \right|^2 .
\end{align}
Note that the probability is obtained by normalizing with the total number of trajectories $\Ntraj$. The probability $\Psub{p}(\nsup{m}, \nsubsup{i}{m^{\!*}} | \nsup{m\minus 1}, \vxsup{m\minus 1} )$ is found by counting the number of idler collapses up until the time $\tsub{m^{\!*}}$ instead of $\tsub{m}$.

With the time dependence of $|\Sin(t)|^2$ fixed, we can introduce the probability that at least one pair is created in bin $m$, $\psup{m}$, as a generalized control setting for the pump. It is calculated using Monte Carlo simulations with the initial condition $\ket{\psi(t_{m\minus 1})} \equal \ket{0, 0}$.

\subsection{\label{app:release model} Releasing}
If we release signal photons in bin $m$, the probability distribution $ \Psub{r}(\nsup{m}, \xsup{m} | \nsup{m\minus 1}, \vxsup{m\minus 1} )$ is also found using~\eqsref{nm and xm given nm1 and xm1}{xm given nm1 vxm1} with $\Isup{m}$ and $\nsubsup{i}{m}$ replaced by $\Ssup{m}$ and $ \nsubsup{s}{m}$, where $\Ssup{m} \equal \xsup{m\minus 1} - \xsup{m}$. Again, the subscript $r$ indicates that we are releasing signal photons in bin $m$.  The probability of different cavity states at $t_m$ is given by the multinomial distribution
\begin{align}\eqlab{Pmultinomial}
\Psub{r}(\nsup{m} ,\nsubsup{s}{m} |\nsup{m\minus 1}, \vxsup{m\minus 1})  = 
\frac{\nsup{m\minus 1}! }{\nsubsup{s}{m}!\hspace{0.5mm}\nsup{m}!\hspace{0.5mm}(\nsup{m\minus 1}\minus\nsubsup{s}{m}\minus\nsup{m})!} \psub{c}^{\nsup{m}} \psub{s}^{\nsubsup{s}{m}} \psub{L}^{\nsup{m\minus 1}-\nsubsup{s}{m}-\nsup{m}}.
\end{align}
The subscripts $\{c,s,L\}$ correspond to the cavity, signal waveguide, and environment loss channel, respectively. The probability that a photon remains in the cavity at time $t$ is $\psub{c}(t)$, the probability that it has coupled into the signal waveguide is $\psub{s}(t)$, and $\psub{L}\equal 1\minus\psub{c}\minus\psub{s}$. The probabilities are found using rate equations for the ensemble average of the number operators
\begin{align} \eqlab{rate eqs}
	\begin{array}{rcl}  
\displaystyle{ \frac{d \expect{\nhatsub{c}(t)} }{dt} } &\!=\!&   - 2\big[\kapsub{s}(t) +\kapL\big] \expect{\nhatsub{c}(t)} \\
&  & \\
\displaystyle{ \frac{d \expect{\nhatsub{s}(t)} }{dt} } &\!=\!&   2\kapsub{s}(t) \expect{\nhatsub{c}(t)}
\end{array} 	\quad \Rightarrow \quad 
	\begin{array}{rcl}  
\psub{c}(t) &\!=\!&   \exp\! \left( \displaystyle{\int_{t_{m\minus 1}}^{t} \!\!\!\!\!\!\!-2\big[\kapsub{s}(t') +\kapL\big] dt'} \right) \\
&  & \\
\psub{s}(t) &\!=\!&   \!\! \displaystyle{\int_{t_{m\minus 1}}^{t} \!\!\!\!\! 2\kapsub{s}(t') \psub{c}(t') dt'} 
\end{array} ,
\end{align}
%
where, e.g. $\expect{\nhatsub{c}(t)} \equal \nsup{m\minus 1} \psub{c}(t) $, and the initial condition $\psub{c}(t_{m\minus 1})\equal 1$ and $\psub{s}(t_{m\minus 1})\equal\psub{L}(t_{m\minus 1})\equal 0$ was used in~\eqref{rate eqs}.  
The distribution $\Psub{r}(\nsup{m} ,\nsubsup{s}{m^{\!*}} |\nsup{m\minus 1}, \vxsup{m\minus 1})$ is found using the intermediary step
\begin{align}\eqlab{Pmultinomial m*}
\Psub{r}(\nsup{m} ,\nsubsup{s}{m^{\!*}} |\nsup{m\minus 1}, \vxsup{m\minus 1}) = 
\!\!\!\!\! \sum_{\nsup{m^{\!*}}\!, \nsubsup{s}{m}} \!\!\!\!\!\! \Psub{r}(\nsup{m} \!,\nsubsup{s}{m} | \nsup{m^{\!*}}\!, \nsubsup{s}{m^{\!*}}\!, \nsup{m\minus 1}\!, \vxsup{m\minus 1}) \Psub{r}(\nsup{m^{\!*}} \!,\nsubsup{s}{m^{\!*}} |\nsup{m\minus 1}\!, \vxsup{m\minus 1}) ,
\end{align}
where the first factor on the RHS only depends on $\nsup{m^{\!*}}$, such that
\begin{align}\eqlab{Pmultinomial m* 2}
\Psub{r}(\nsup{m} ,\nsubsup{s}{m^{\!*}} |\nsup{m\minus 1}, \vxsup{m\minus 1}) = 
\!\!\!\! \sum_{\nsup{m^{\!*}}\!, \nsubsup{s}{m}} \!\!\!\!\!\! \Psub{r}(\nsup{m} \!,\nsubsup{s}{m} | \nsup{m^{\!*}}) \Psub{r}(\nsup{m^{\!*}} \!,\nsubsup{s}{m^{\!*}} |\nsup{m\minus 1}\!, \vxsup{m\minus 1}). 
\end{align}
The distribution $\Psub{r}(\nsup{m} \!,\nsubsup{s}{m} | \nsup{m^{\!*}})$ must be evaluated using~\eqsref{Pmultinomial}{rate eqs} with the chosen temporal evolution of $\kapsub{s}(t)$. 

The faster signal photons couple into the waveguide, the smaller is the probability that they will be lost to the environment. If 
$\psub{c}(t)$ drop instantly from 1 at $t_{m\minus 1}$ to its final value then such loss is avoided. This corresponds to $\kapsub{s}(t)$ being proportional to a Dirac delta distribution centered at $t_{m\minus 1}$. However, the electrical signal that controls the signal filter has some finite temporal width. We assume its shape is Gaussian and that the temporal shape of the coupling is
\begin{align}\eqlab{kapsD}
\kapsub{s}(t) \propto  \int_{t_{m\minus 1}}^{t} \!\!\!\! e^{-\kappsiS (t-t')} e^{-(t'-\tsub{r})^2/\tau_r^2 } dt' .
\end{align} 
where $1/\kappsiS$ is a response time, $\tsub{r}$ is adjusted such that $\kapsub{s}(t_{m\minus 1})$ is at least a thousand times smaller than its peak value and the width, $\tau_r$, is constrained by the condition $\kapsub{s}(t)\!\leq\! \kapsub{i}$ seen in~\figref{gate prop 1}b. With the shape of $\kapsub{s}(t)$ given by~\eqref{kapsD}, the release control setting is completely determined by the value $\psubsup{c}{m}\equiv \psub{c}(\tsub{m})$ calculated from~\eqref{rate eqs}.

\section{\label{sec:protocol} Driving protocol}
The driving protocol relates information from photon detections to control actions.
\figref{flow chart} depicts which actions are taken in bin $m\plus 1$ depending on the updated probability distribution in bin $m$,  $P(\nsup{m} | \vxsup{m^{\!*}} )$. 
\begin{figure}[!h] 
  \centering
  \includegraphics[width=11cm]{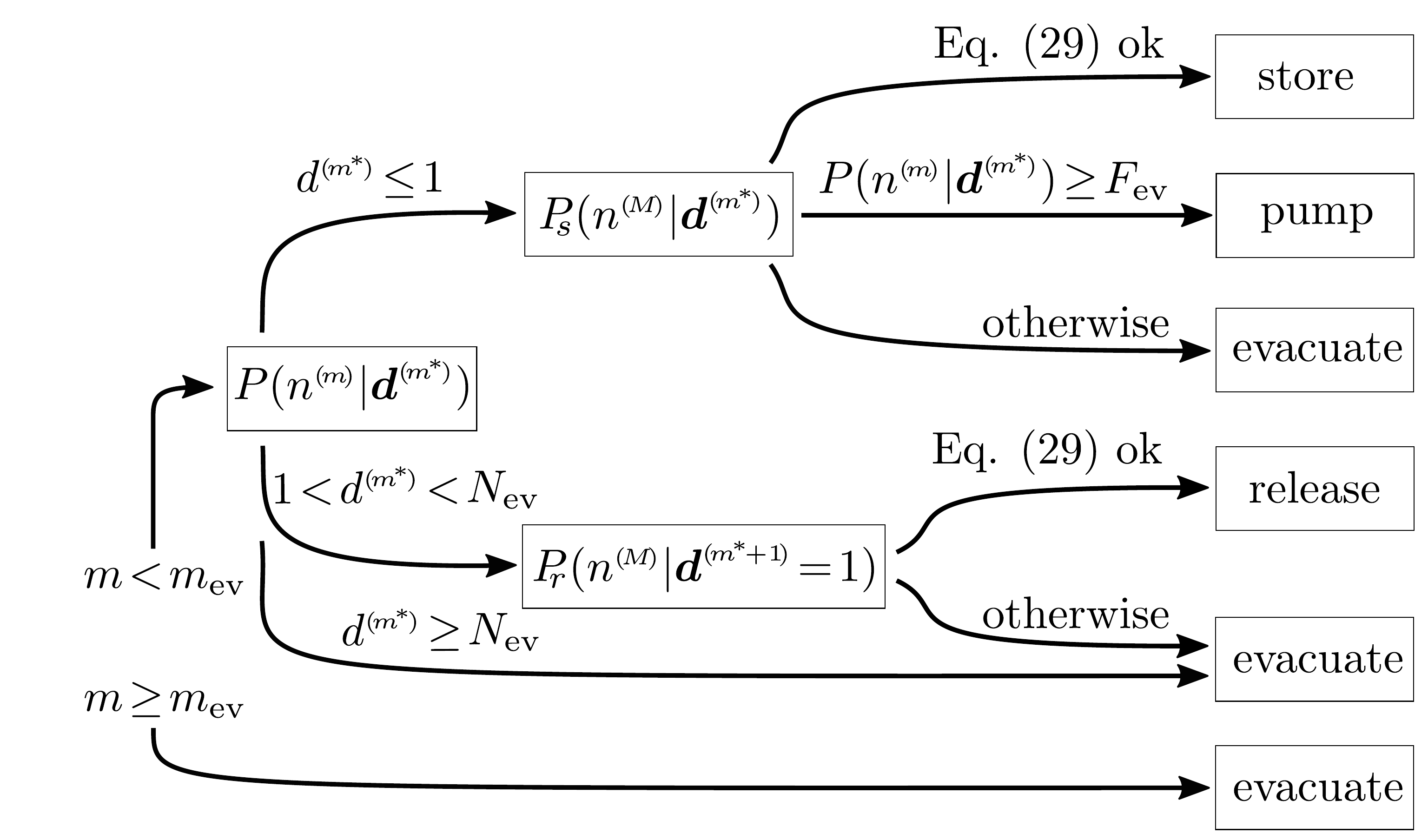}
 \caption{Diagram illustrating the driving protocol.}
\figlab{flow chart}
\end{figure}
In some parameter regimes, it turns out to be advantageous to evacuate the cavity after a certain number of bins, $\mevac$, irrespective of the state estimation fidelity. If $m\!<\!\mevac$, the protocol depends on $\vxsup{m^{\!*}}$ in the following way:  
For $\xsup{m^{\!*}}\!\leq\! 1$, we evaluate the distribution  at the emission time in case the state is stored until then, $\Psub{s}(\nsup{M} | \vxsup{m^{\!*}} )$. If~\eqref{criteria} is fulfilled, the state is stored. If not, the next bin is pumped if the fidelity of our current state estimation is larger than the control parameter $\Fevac$ and otherwise evacuated.  
For $ 1\!<\!\xsup{m^{\!*}}\!<\! \Nevac$, we calculate the distribution $\Psub{r}(\nsup{M} | \vxsup{m^{\!*}\plus 1}\!\equal\! 1 )$ corresponding to a release in the next bin followed by storage until the emission time. Again, if the threshold conditions are met by evaluating~\eqref{criteria}, we release in the following bin and evacuate otherwise. $\Nevac$ is a control parameter that primarily serves to reduce the optimization space. However, it also allows us to eliminate release from the protocol by choosing $\Nevac\equal 2$.

In the following sections we provide more details about the protocol in case of pumping and releasing, respectively.

\subsection{\label{app:pump protocol} Pumping protocol}
When pumping in bin $m$, we assume that it is advantageous to keep the pump power below a level, which for the initial condition $\ket{\psi(\tsub{m\minus 1})} \equal \ket{0,0}$ results in a distribution $\Psub{s}(\nsup{M}\equal 1 | \vxsup{m^{\!*}})$ that obeys the requirements in~\eqref{criteria}. It is given by
\begin{align} \eqlab{herald eff m to e}
\Psub{s}(\nsup{M} | \vxsup{m^{\!*}} ) =  \!\!\!\!\!\!\!\! \sum_{\nsup{m} = \nsup{M}}^\infty \!\!\!\!\!\!\! \Psub{s}(\nsup{M} | \nsup{m}, \vxsup{m^{\!*}}) P(\nsup{m} | \vxsup{m^{\!*}}),
\end{align}
where the subscript $s$ indicates storage until $\tM$. The probability that there are $\nsup{M}$ signal photons in the cavity at $\tsub{M}$ given there was $\nsup{m}$ at $\tsub{m}$ and the detection sequence was $\vxsup{m^{\!*}}$ is given by~\eqref{Pmultinomial} with $\psub{s} \equal \nsub{s} \equal 0$
\begin{align}\eqlab{Pbinom}
\Psub{s}(\nsup{M} | \nsup{m}, \vxsup{m})  = \frac{\nsup{m}! \hspace{0.5mm} \psub{c}^{\nsup{M}} (1\minus\psub{c})^{\nsup{m} \!- \nsup{M}} }{\nsup{M}!\hspace{0.5mm}(\nsup{m}\minus\nsup{M})!},
\end{align}
where $\psub{c} \equal \exp[-2\kapL(\tM\minus t_m)]$ is the probability that a photon remains in the cavity until $\tM$.

~\figref{pump limits} shows how the relevant properties of the distribution $P(\nsup{m} | \xsup{m^{\!*}})$ vary as a function of the pump control setting, $\psup{m}$, and bin duration, $\taubin$.  
\begin{figure}[h]
  \centering
  \includegraphics[height=5.5cm]{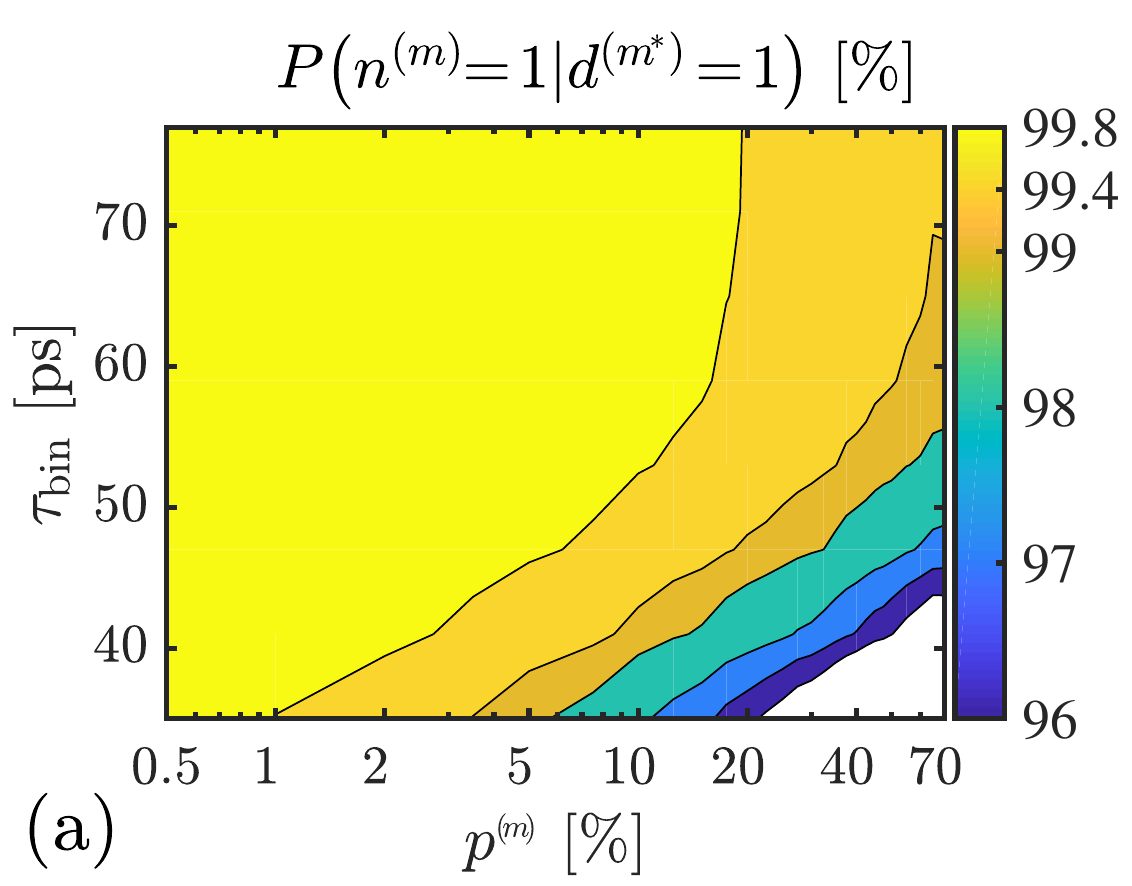}
	\hspace{5mm}
  \includegraphics[height=5.5cm]{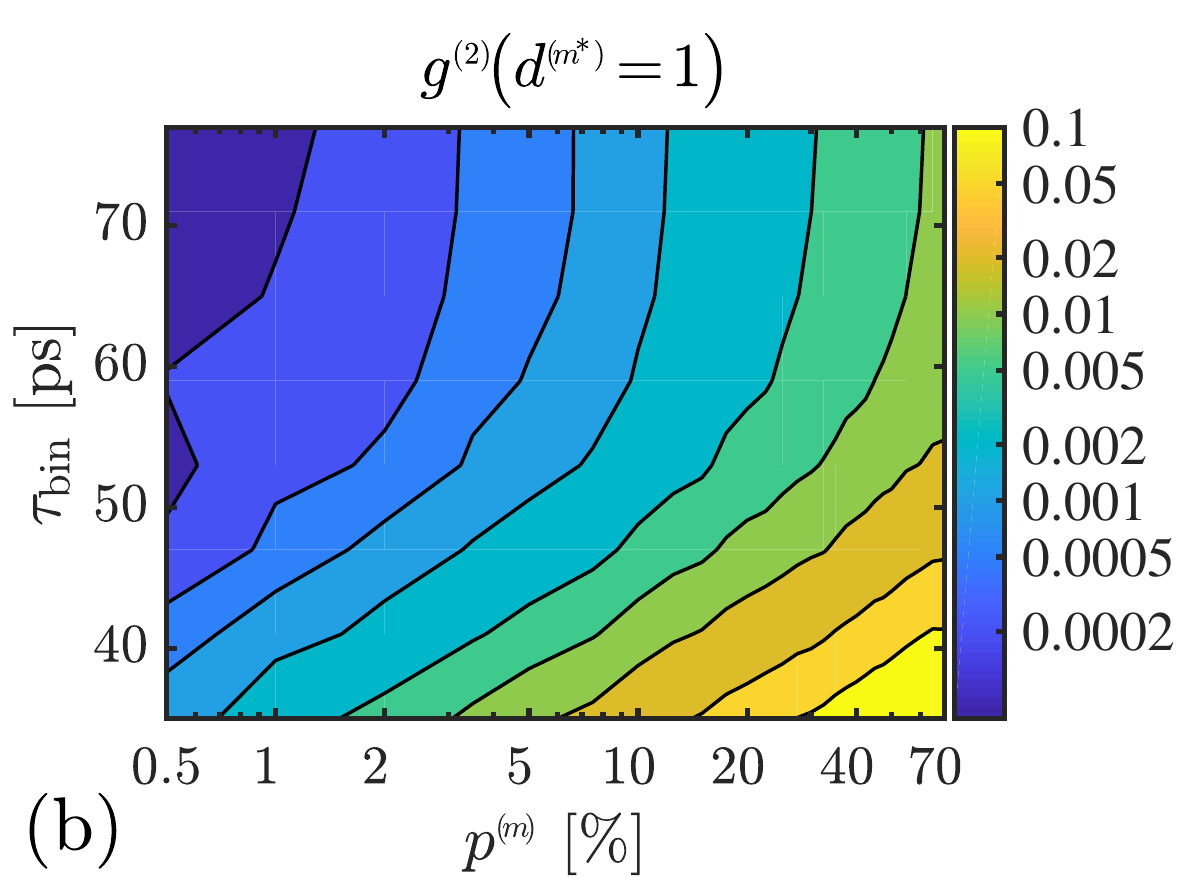} 
 \caption{ Conditional probability (a) 
 and second-order correlation (b) 
 after pumping in bin $m$ as a function of the pair creation probability and bin duration. The parameters are: $\Qintr\equal $ 200$\!\times\!10^6$, $\Qsub{p}\equal \Qsub{i} \equal 6667$, $\Delta_s\equal -\Delta_i \equal 20\kapsub{p}$, $\eta\equal 0.996$, $\tD\equal 12$ps,  $\ket{\psi(t_{m\minus 1})} \equal \ket{0, 0}$. }
\figlab{pump limits}
\end{figure} 
For small $\taubin$, our state estimation fidelity is degraded by the finite probability that idler photons remain inside the cavity. As the bin duration is increased, the fidelity eventually becomes dominated by the detection efficiency and the contours become vertical. This suggests that there must be an optimum bin duration, because increasing it allows larger pump power while reducing the total number of bins. The properties illustrated in~\figref{pump limits} apply to any bin, but since $P_s(\nsup{M} | \vxsup{m^{\!*}})$ depends on $M\minus m$, the maximum control setting for the pump must be evaluated for each bin. 

\subsection{\label{app:release protocol} Release protocol}

To test whether releasing in bin $m$ can lead to a successful state creation, the success conditions in~\eqref{criteria} can be evaluated using~\eqref{herald eff m to e} for all possible outcomes $\{\nsup{m},~\xsup{m^{\!*}}\}$ with a fixed control setting $\psubsup{c}{m}$. If~\eqref{criteria} can be fulfilled, the control setting is adjusted to maximize the success probability. Since we are only interested in the high detection efficiency regime, we assume that only $\xsup{m^{\!*}}\equal 1$ provides a sufficiently large state fidelity and therefore only use the outcomes $\{\nsup{m}, 1\}$ to optimize the control setting. If~\eqref{criteria} can not be met, the cavity is evacuated. Determining whether it is advantageous to release or evacuate requires an evaluation of the success probability for a very large number of outcome scenarios. We therefore simplify the release protocol by always evacuating if $\xsup{m^{\!*}\minus 1} \!\geq \! \Nevac$.

While the optimization procedure described above is done numerically, insight into the maximum success probability of the release process may be gained by considering the probability that $\nsup{m\plus 1}\equal 1$ after releasing in bin $m\plus 1$. It is given by~\eqref{Pmultinomial}
\begin{align} 
P(1 |\nsup{m}\!, \vxsubsup{r}{m^{\!*}})  &= \!\!\!\sum_{\nsubsup{s}{m} = 0} ^{\nsup{m}\minus 1} \! \frac{\nsup{m}! \psub{c} \psub{s}^{\nsubsup{s}{m}} \psub{L}^{\nsup{m} \minus 1\minus\nsubsup{s}{m}}}{\nsubsup{s}{m}!(\nsup{m}\minus 1\minus \nsubsup{s}{m} )!}  =  \nsup{m} \psub{c} \!\!\sum_{\nsubsup{s}{m} = 0} ^{\nsup{m}\minus 1} \!\!\left( \!\!\!\begin{array}{c}
\nsup{m}\minus 1\\
\nsubsup{s}{m}
\end{array} \!\!\!\right) \psub{s}^{\nsubsup{s}{m}} \psub{L}^{\nsup{m}-1-\nsubsup{s}{m}}  \nn\\
 &= \nsup{m} \psub{c} (\psub{s} \plus \psub{L})^{\nsup{m} \minus 1} = \nsup{m} \psub{c} (1 \minus \psub{c})^{\nsup{m}\minus 1} . \eqlab{P1max release}
\end{align}
The maximum is found when $\nsup{m} \psub{c} \equal 1$ and equals $1/2$ for $\nsup{m}\equal 2$ and $4/9$ for $\nsup{m}\equal 3$. For perfect detectors this illustrates that the probability of reaching a single photon state after initially creating two or three pairs is rather good.

\subsection{\label{app:optimize protocol} Optimization of protocol}
To optimize the protocol, we start by assuming there is only a single time bin available and iteratively increase $M$ while keeping track of the success probability.
When evacuating in bin $m$, the optimum strategy for the remaining $M\minus m$ bins is known from a previous iteration. The contribution to the overall success probability is $P (\vxsup{m^{\!*}\minus 1}) \PmcS(M\minus m)$, where $\vxsup{m^{\!*}\minus 1}$ is the detection sequence leading us to evacuate in bin $m$. For instance, for $M\equal 3$ there is a possibility that $\xsup{1^{\!*}}$ causes us to evacuate the cavity in bin 2. In this scenario, we treat the last bin as the $M\equal 1$ case because the initial condition of bin 3 will be $\ket{\psi(\tsub{2})}\equal \ket{0,0}$ after evacuation. The protocol parameters for the last bin are then set as the optimum parameters found for $M\equal 1$. The iteration continues until~\eqref{criteria} can not be met for larger $M$. An upper bound on $M$ is found by considering that $\Fth$ must be larger than the probability of loosing a signal photon within $M$ bins, $\exp[-2\kapL M\taubin]$.
A fixed pump sequence, $\vpsup{M} \equiv \{\psup{1}, \psup{2}, \ldots, \psup{M\minus 1}, \psup{M} \}$, is used for all detection sequences $\vxsup{M}$ unless an evacuation occurs in bin $m\!<\!M$.
\begin{figure}[!htb]
  \centering
  \includegraphics[height=5.2cm]{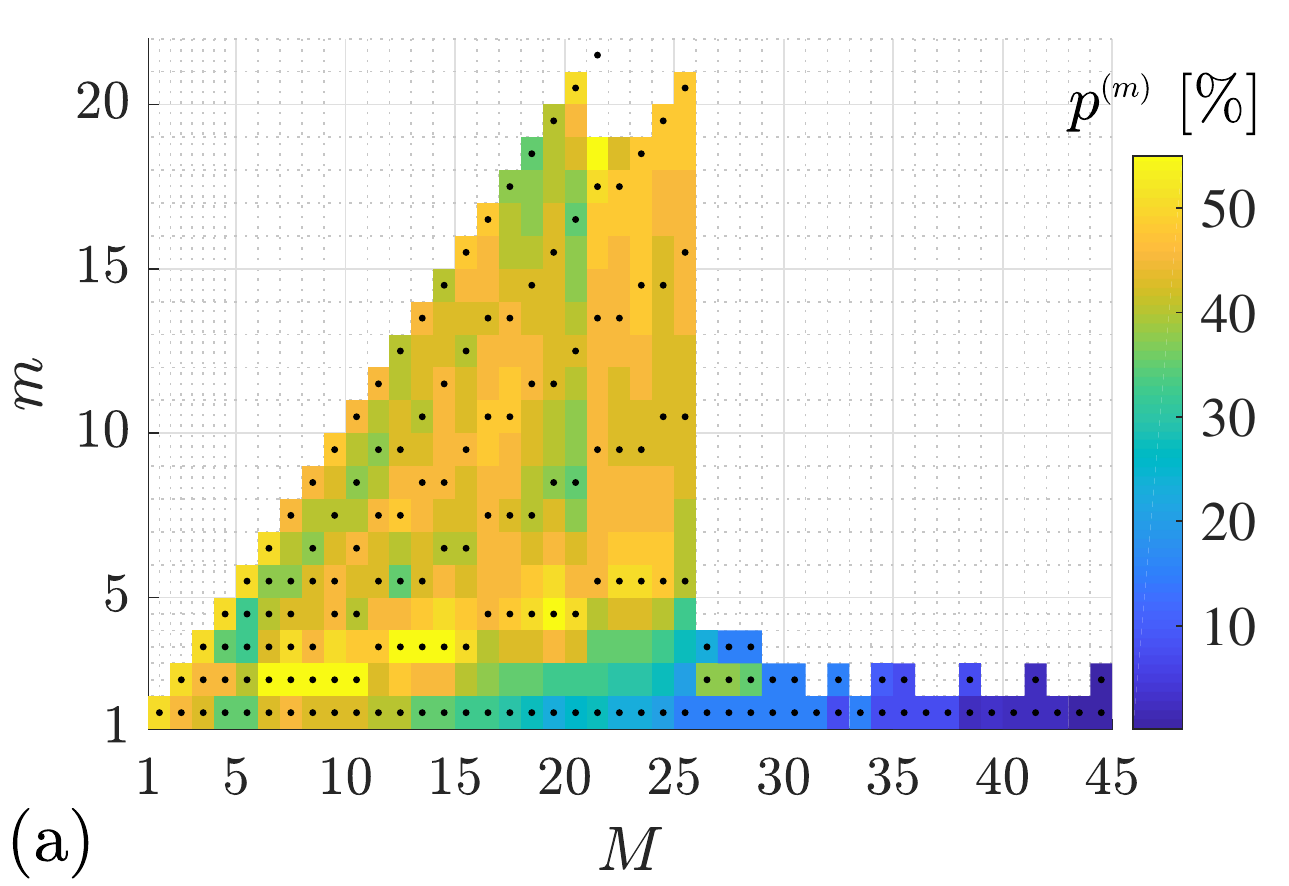}
  \hspace{4mm}
  \includegraphics[height=5.2cm]{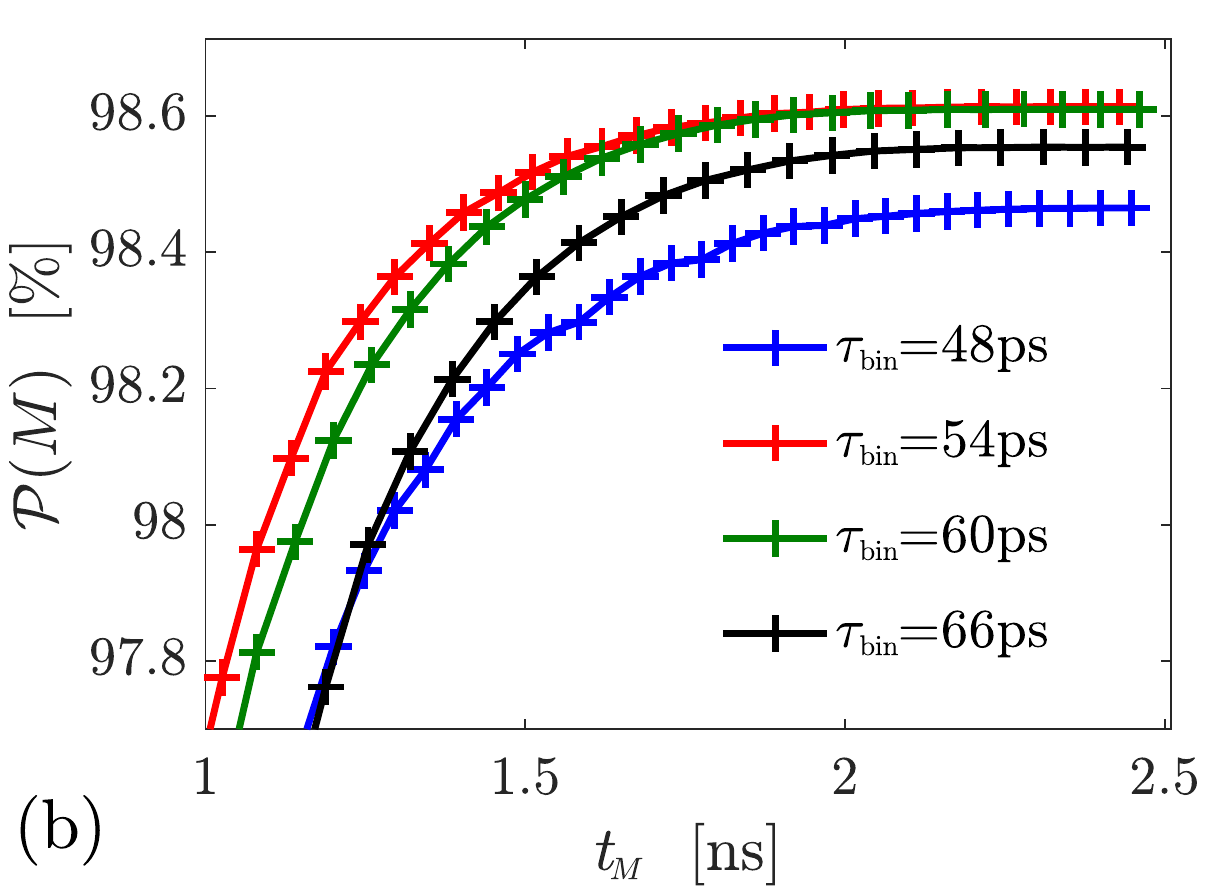}
 \caption{(a) Optimized probabilities of creating at least one photon pair as a function of $m$ for different lengths of the emission cycle, $M$. The black dots indicate bins used for linear interpolation.  
(b) Success probability as a function of $\tsub{M}$ for different bin duration. The $+$ markers indicate time-bin separations. The parameters are: $\Qintr\equal 200\!\times\!10^6$, $\eta\equal 0.996$, $\Fth \equal 0.985$, $\gllth\equal 1.0$, $\Nevac\equal 3$.}
\figlab{Psucc optim pump}
\end{figure}
The single bin probability distribution in~\eqref{PsCiD} has been found using Monte Carlo simulations as a function of $\psup{m}$ for the discrete set of values $\psup{m}\in [0.1,~0.2,~0.5,~1,~2,~3.5,~5-50,~55,~60,~65,~70]$, where the step-size between 5 and 50 is 2.5 and all numbers are given in \%.

All the control parameters $\Nevac,~\Fevac$, $\mevac$, and $\vpsup{M}$ are optimized for each $M$ to maximize the success probability. Additionally, $\psubsup{c}{m}$ is optimized for each bin and $\taubin$ is optimized for the entire emission cycle.
\figref{Psucc optim pump}a shows an example of the optimized pump sequences.   
For each $M$ on the horizontal axis, the vertical axis plots the sequence $\vpsup{m}$ with colors indicating the value of $\psup{m}$. For $M>20$, the cavity is always evacuated in a bin $m$, for which $m<M$. For $M>25$, the optimum strategy is to evacuate at $m\equal \mevac$ and the number of colored bins is given by $\mevac\minus 1$. As the number of possible pump sequences increases exponentially with $M$, we optimize using a small sample of bins and use linear interpolation to find $\psup{m}$ for all bins. The black dots indicate which bins are used in the interpolation. 
\figref{Psucc optim pump}b shows how the optimized success probability depends on the bin duration.
The optimum choice for $\taubin$ is seen to correspond to where the contour lines in~\figref{pump limits}a start to become vertical. It seems reasonable that longer bins are sub-optimal because~\figref{pump limits} suggests that no benefit from increased pump power is possible.

\section{\label{sec:results} Simulation Results}
The system performance is evaluated by fixing the static coupling-$Q$ of the idler and pump cavity modes at $Q_n\equal \omega_n/2\kappa_n \equal 6667$ $(n\equal i,p)$ and optimizing the success probability for loss rates corresponding to intrinsic quality factors of 40, 80, and 200 million.
\begin{figure}[!htb]
  \centering
  \includegraphics[height=5.2cm]{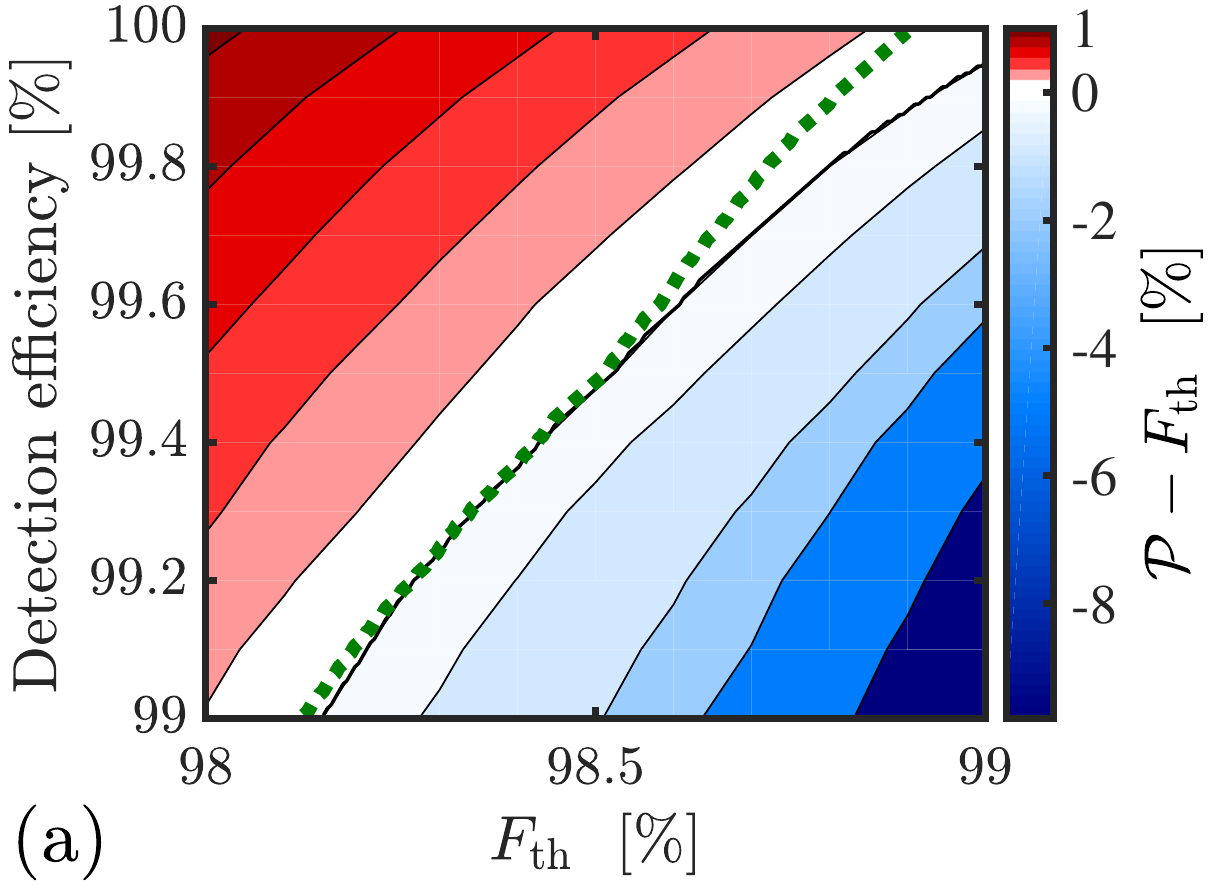}
  \hspace{6mm}
  \includegraphics[height=5.2cm]{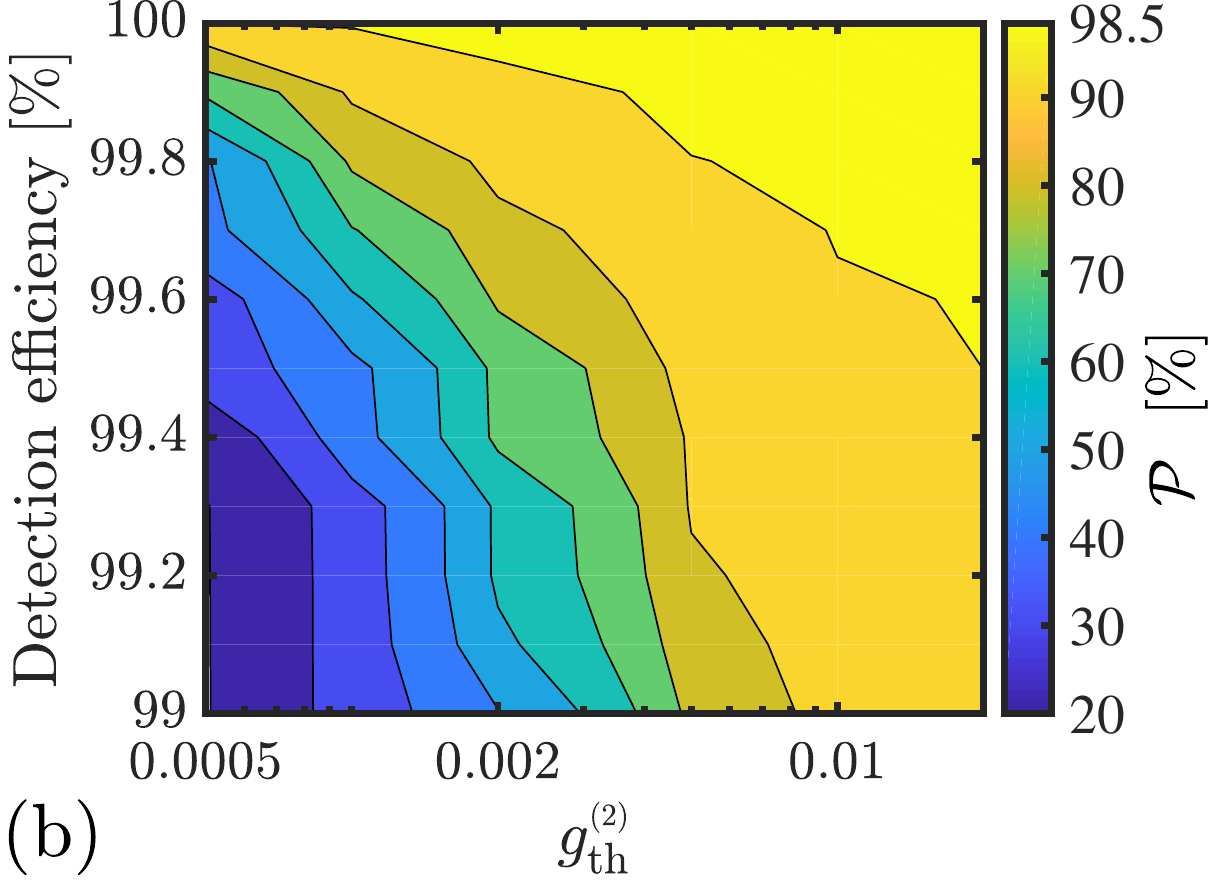}
 \caption{(a) Difference between the maximum success probability and state fidelity threshold as a function of $\Fth$ and detection efficiency. The parameters are: $\Qintr\equal 20\mytimes 10^7$,  $\DelI\equal -\DelS \equal 20\kapsub{i}$, $\tD\equal 12$ps, $1/\kappsiS\equal 3$ps,  $\gllth\equal 1.0$, and $\Nevac \equal 3$. The dotted green curve shows the contour corresponding to $\PmcS\equal \Fth$ for $\Nevac\equal 2$. (b) Success probability as a function of $\gllth$ and $\eta$ for $\Fth\equal 98.5\%$.}
\figlab{results 1}
\end{figure}
\figref{results 1}a shows the trade-space between the state fidelity threshold and success probability at different detection efficiency for $\Qintr\equal$ 200$\!\times\!10^6$ and $\gllth \equal 1.0$ (this large value of $\gllth$ ensures that it does not limit the success probability). Remarkably, a success probability and state fidelity of 99\%  is achievable for $\eta$ just below unity. Alternatively, a 99\% state fidelity is achieved with a success probability of $\PmcS\equal 89.2\%$ if $\eta\equal 0.99$. Note that $\PmcS\!>\!\Fth$ is possible because some detection sequences result in a state fidelity larger than the threshold. \figref{results 1}b shows the effect of reducing the second order correlation threshold for $\Fth\equal 98.5\%$. It illustrates the reduction in success probability when requiring a lower multi-photon contamination level.\\
\begin{figure}[!htb]
  \centering
  \includegraphics[height=5.2cm]{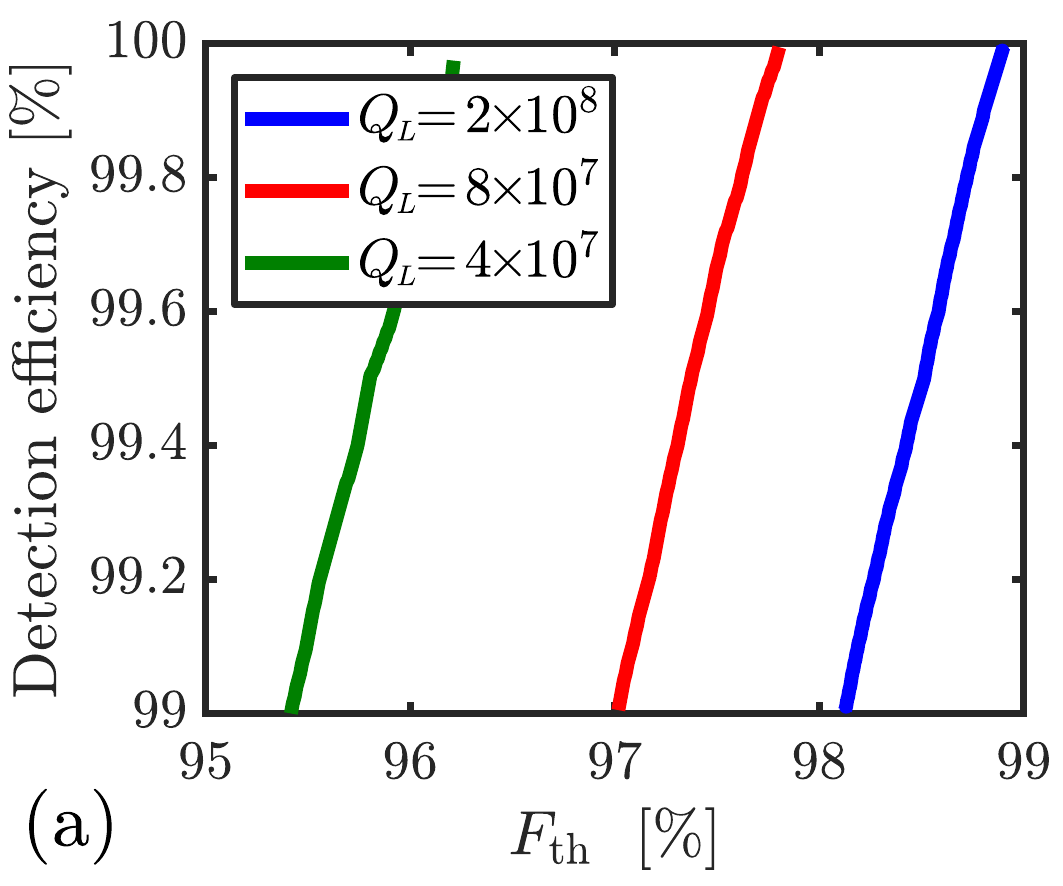}
  \hspace{10mm}
  \includegraphics[height=5.2cm]{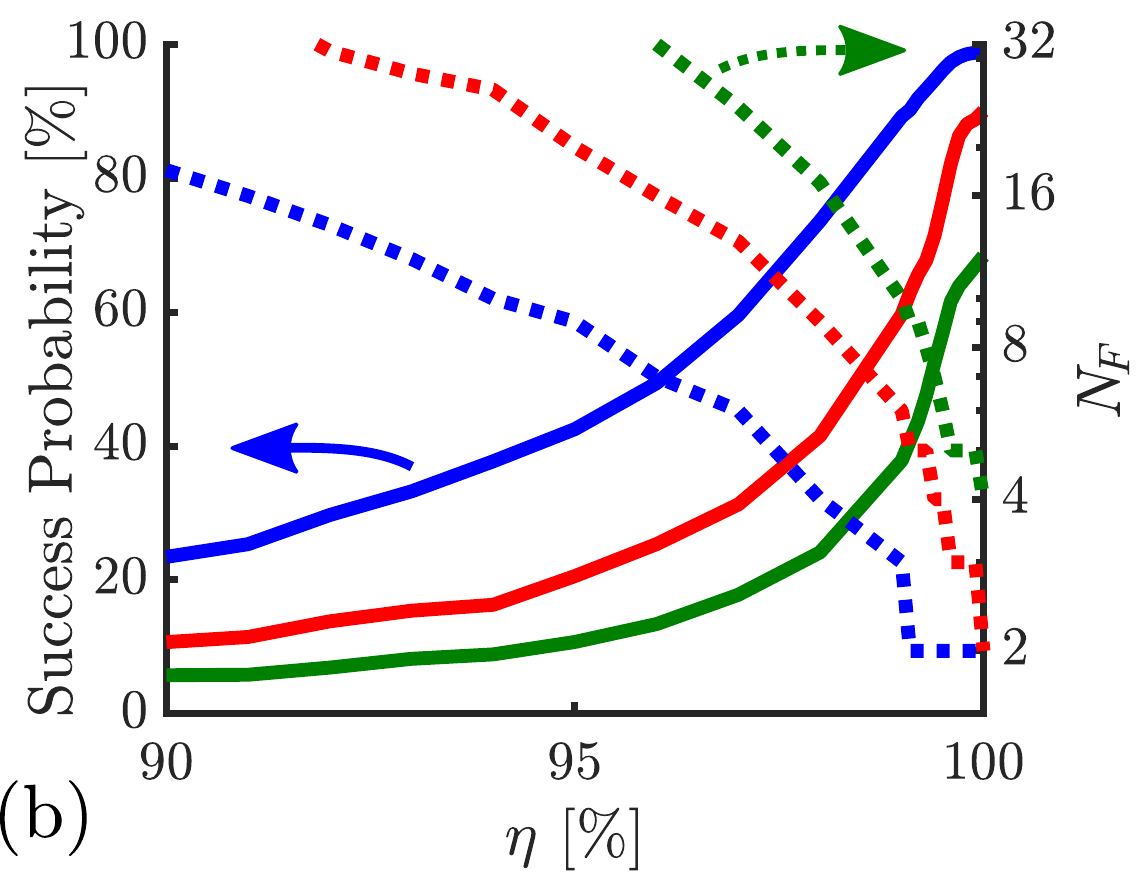}
 \caption{(a) Contour-lines $\PmcS\equal \Fth$ for three different loss rates. (b) Left axis: Success probability as a function of detection efficiency for $\Fth\equal 99\%$ (solid lines). Right axis: Number of parallel frequency modes necessary to achieve an efficiency of $\mcPtot\geq 99\%$ (dotted lines). The parameters and legends are the same as in~\figref{results 1} except $\Nevac \equal 2$.}
\figlab{results 2}
\end{figure}

We investigated the detrimental effect of increasing the loss rate and how performance can be restored using frequency multiplexing.~\figref{results 2}a shows the contour-lines $\PmcS\equal \Fth$ for three different values of $\Qintr$. In~\figref{results 2}b we plot the success probability with $\Fth\equal 99\%$ as a function of detection efficiency. \figref{results 2}b also shows the necessary number of parallel frequency modes to achieve a total success probability of $\mcPtot \equal 99\%$. For instance, reducing $\Qintr$ from 200 to 40 million only requires 4 frequency modes to restore $\mcPtot$ for $\eta\equal 1$.

\section{\label{sec:discussion} Discussion}
Our analysis shows that heralded single photon sources should be possible by on-chip multiplexing with near-unity purity. However, the device requirements are stringent - especially on the detector and feed-forward. 

State of the art demonstrations of chip-integrated resonators~\cite{Biberman2012, Yang2017, Ji2017} have reached quality factors exceeding 200 million, showing that the required intrinsic quality factors are within reach.
For our protocol, the storage-ring round-trip time should be much shorter than the pump pulse duration. Practically, this means that the storage ring should be less than $\sim \!100\mu$m in circumference. Larger devices with longer pump pulses can be used at the cost of reducing the number of available time bins.

The only detector technology that is currently able to approach the performance requirements are superconducting nanowire single photon detectors (SNSPDs)~\cite{Divochiy2008, Marsili2013, Schuck2013, Najafi2015, Schelew2015}. Electronic logic~\cite{McCaughan2014} and electro-optic switching~\cite{Gehl2017} have been demonstrated at cryogenic temperatures, which are necessary when using SNSPDs and assuming a control set-time, $\tD$, on the order of 10ps.

As seen from~\figref{results 1}a, the performance is not significantly reduced if the release step is omitted $(\Nevac\equal 2)$.
This means that the tunable signal filter only needs to be low-loss in its closed state since loss during evacuation is irrelevant. In comparison, switches for spatial multiplexing must be low-loss in both states. The tunable output filter must, however, be low-loss in both settings. Since the output filter is only used for emission (once per emission cycle), one might consider using two different physical mechanisms to tune $\Delta\psi_s$ and $\Delta\psi_o$, such as carrier dispersion and heating. Developments in reducing the thermal response time in nanophotonic structures~\cite{Khurgin2015} could be a path towards high-speed low-loss switching.

Using cavity modes with different coupling rates for the idler, pump, and signal has been shown to enable signal-idler states with joint spectral amplitudes that are almost completely separable (implying high purity signal states)~\cite{Vernon2017a, Tison2017}. The main challenges for creating indistinguishable photons with our proposed architecture is stabilization of the high $Q$ resonance and repeatability of the output filter opening. The purity and temporal shaping of photons emitted from our proposed architecture will be studied in more detail in future work. 

In conclusion, near-unity efficiency and photon purity is achieved by Bayesian inference, based on known system parameters and photon detections; similar Bayesian state-estimation should also be useful for improving bulk-optics multiplexed sources~\cite{Glebov2013, Kaneda2015a} and relative-multiplexing schemes~\cite{Gimeno-segovia2017}.
This proposal of near-perfect on-chip single photon sources substantiates the feasibility of quantum technologies that require the production of large-scale photonic quantum states, such as optical quantum repeater networks,  precision measurements, and quantum computing systems.


\section*{Acknowledgements}
MP and DE acknowledge support from AFOSR MURI for Optimal Measurements for Scalable Quantum Technologies (FA9550-14-1-0052) and the Air Force Research Laboratory RITA program (FA8750-14-2-0120). MP acknowledges support from DARPA project Scalable Engineering of Quantum Optical Information Processing Architectures (SEQUOIA), under US Army contract number W31P4Q-15-C-0045 . MH acknowledges support from the Danish Council for Independent Research (DFF1325-00144) and the Velux Foundations.

\section*{References}


\begin{thebibliography}{10}

\bibitem{Kok2007}
P.~Kok, K.~Nemoto, T.~C. Ralph, J.~P. Dowling, and G.~J. Milburn, ``{Linear
  optical quantum computing with photonic qubits},'' {\em Reviews of Modern
  Physics}, vol.~79, pp.~135--174, 1 2007.

\bibitem{Somaschi2015}
N.~Somaschi, V.~Giesz, L.~De~Santis, J.~C. Loredo, M.~P. Almeida, G.~Hornecker,
  S.~L. Portalupi, T.~Grange, C.~Anton, J.~Demory, C.~Gomez, I.~Sagnes,
  N.~D.~L. Kimura, A.~Lemaitre, A.~Auffeves, A.~G. White, L.~Lanco, and
  P.~Senellart, ``{Near optimal single photon sources in the solid state},''
  {\em Nature Photonics}, vol.~10, no.~2, 2016.

\bibitem{Ding2016}
X.~Ding, Y.~He, Z.~C. Duan, N.~Gregersen, M.~C. Chen, S.~Unsleber, S.~Maier,
  C.~Schneider, M.~Kamp, S.~H{\"{o}}fling, C.-Y. Lu, and J.-W. Pan,
  ``{On-Demand Single Photons with High Extraction Efficiency and Near-Unity
  Indistinguishability from a Resonantly Driven Quantum Dot in a
  Micropillar},'' {\em Physical Review Letters}, vol.~116, no.~January,
  pp.~1--6, 2016.

\bibitem{Aharonovich2016}
I.~Aharonovich, D.~Englund, and M.~Toth, ``{Solid-state single-photon
  emitters},'' {\em Nature Photonics}, vol.~10, no.~10, pp.~631--641, 2016.

\bibitem{Collins2013}
M.~J. Collins, C.~Xiong, I.~H. Rey, T.~D. Vo, J.~He, S.~Shahnia, C.~Reardon,
  T.~F. Krauss, M.~J. Steel, a.~S. Clark, and B.~J. Eggleton, ``{Integrated
  spatial multiplexing of heralded single-photon sources.},'' {\em Nature
  communications}, vol.~4, no.~May, p.~2582, 2013.

\bibitem{Kaneda2015a}
F.~Kaneda, B.~G. Christensen, J.~J. Wong, H.~S. Park, K.~T. McCusker, and P.~G.
  Kwiat, ``{Time-multiplexed heralded single-photon source},'' {\em Optica},
  vol.~2, no.~12, p.~1010, 2015.

\bibitem{Glebov2013}
B.~L. Glebov, J.~Fan, and a.~Migdall, ``{Deterministic generation of single
  photons via multiplexing repetitive parametric downconversions},'' {\em
  Applied Physics Letters}, vol.~103, no.~3, p.~031115, 2013.

\bibitem{Joshi2017}
C.~Joshi, A.~Farsi, S.~Clemmen, S.~Ramelow, and A.~L. Gaeta, ``{Frequency
  Multiplexing for Quasi-Deterministic Heralded Single-Photon Sources},'' {\em
  ArXiv}, 2017.

\bibitem{pant2017rate}
M.~Pant, H.~Krovi, D.~Englund, and S.~Guha, ``Rate-distance tradeoff and
  resource costs for all-optical quantum repeaters,'' {\em Physical Review A},
  vol.~95, no.~1, p.~012304, 2017.

\bibitem{Giovannetti2011}
V.~Giovannetti, S.~Lloyd, and L.~Maccone, ``{Advances in quantum metrology},''
  {\em Nature Photonics}, vol.~5, no.~4, pp.~222--229, 2011.

\bibitem{li2015}
Y.~Li, P.~C. Humphreys, G.~J. Mendoza, and S.~C. Benjamin, ``Resource costs for
  fault-tolerant linear optical quantum computing,'' {\em Physical Review X},
  vol.~5, no.~4, p.~041007, 2015.

\bibitem{pant2017percolation}
M.~Pant, D.~Towsley, D.~Englund, and S.~Guha, ``Percolation thresholds for
  photonic quantum computing,'' {\em arXiv preprint arXiv:1701.03775}, 2017.

\bibitem{Barbarossa1995}
G.~Barbarossa, A.~M. Matteo, and M.~N. Armenise, ``{Theoretical Analysis of
  Triple-Coupler Ring-Based Optical Guided-Wave Resonator},'' {\em Journal of
  Lightwave Technology}, vol.~13, no.~2, pp.~148--157, 1995.

\bibitem{Li2016}
Q.~Li, M.~Davanco, and K.~Srinivasan, ``{Efficient and low-noise
  single-photon-level frequency conversion interfaces using silicon
  nanophotonics},'' {\em Nature Photonics}, vol.~10, no.~June, pp.~406--415,
  2016.

\bibitem{McKinstrie2005}
C.~J. McKinstrie, J.~D. Harvey, S.~Radic, and M.~G. Raymer, ``{Translation of
  quantum states by four-wave mixing in fibers},'' {\em Optics Express},
  vol.~13, no.~22, p.~9131, 2005.

\bibitem{Vernon2016}
Z.~Vernon, M.~Liscidini, and J.~E. Sipe, ``{Quantum frequency conversion and
  strong coupling of photonic modes using four-wave mixing in integrated
  microresonators},'' {\em Physical Review A - Atomic, Molecular, and Optical
  Physics}, vol.~94, no.~2, pp.~1--14, 2016.

\bibitem{Huang2013}
Y.-P. Huang, V.~Velev, and P.~Kumar, ``{Quantum frequency conversion in
  nonlinear microcavities.},'' {\em Optics Letters}, vol.~38, pp.~2119--21, jun
  2013.

\bibitem{Li2016a}
A.~Li, T.~Chen, Y.~Zhou, and X.~Wang, ``{On-demand single-photon sources via
  quantum blockade and applications in decoy-state quantum key distribution},''
  {\em Optics Letters}, vol.~41, no.~9, pp.~2--5, 2016.

\bibitem{Johansson2013}
J.~R. Johansson, P.~D. Nation, and F.~Nori, ``{QuTiP 2: A Python framework for
  the dynamics of open quantum systems},'' {\em Computer Physics
  Communications}, vol.~184, no.~4, pp.~1234--1240, 2013.

\bibitem{Haus1991}
H.~Haus and W.~Huang, ``{Coupled-mode theory},'' {\em Proceedings of the IEEE},
  vol.~79, no.~10, pp.~1505--1518, 1991.

\bibitem{Biberman2012}
A.~Biberman, M.~J. Shaw, E.~Timurdogan, J.~B. Wright, and M.~R. Watts,
  ``{Ultralow-loss silicon ring resonators},'' {\em Optics Letters}, vol.~37,
  no.~20, pp.~4236--4238, 2012.

\bibitem{Yang2017}
K.~Y. Yang, D.~Y. Oh, S.~H. Lee, Q.-F. Yang, X.~Yi, and K.~Vahala,
  ``{Integrated Ultra-High-Q Optical Resonator},'' {\em ArXiv}, pp.~1--5, 2017.

\bibitem{Ji2017}
X.~Ji, F.~A.~S. Barbosa, S.~P. Roberts, A.~Dutt, J.~Cardenas, Y.~Okawachi,
  A.~Bryant, A.~L. Gaeta, and M.~Lipson, ``{Ultra-low-loss on-chip resonators
  with sub-milliwatt parametric oscillation threshold},'' {\em Optica}, vol.~4,
  no.~6, p.~619, 2017.

\bibitem{Divochiy2008}
A.~Divochiy, F.~Marsili, D.~Bitauld, A.~Gaggero, R.~Leoni, F.~Mattioli,
  A.~Korneev, V.~Seleznev, N.~Kaurova, O.~Minaeva, G.~Gol'tsman, K.~G.
  Lagoudakis, M.~Benkhaoul, F.~L{\'{e}}vy, and A.~Fiore, ``{Superconducting
  nanowire photon-number-resolving detector at telecommunication
  wavelengths},'' {\em Nature Photonics}, vol.~2, no.~5, pp.~302--306, 2008.

\bibitem{Marsili2013}
F.~Marsili, V.~B. Verma, J.~a. Stern, S.~Harrington, a.~E. Lita, T.~Gerrits,
  I.~Vayshenker, B.~Baek, M.~D. Shaw, R.~P. Mirin, and S.~W. Nam, ``{Detecting
  Single Infrared Photons with 93{\%} System Efficiency},'' {\em Nature
  Photonics}, vol.~7, no.~February, pp.~210--214, 2013.

\bibitem{Schuck2013}
C.~Schuck, W.~H.~P. Pernice, and H.~X. Tang, ``{Waveguide integrated low noise
  NbTiN nanowire single-photon detectors with milli-Hz dark count rate},'' {\em
  Scientific Reports}, vol.~3, p.~1893, 2013.

\bibitem{Najafi2015}
F.~Najafi, J.~Mower, N.~C. Harris, F.~Bellei, A.~Dane, C.~Lee, X.~Hu,
  P.~Kharel, F.~Marsili, S.~Assefa, K.~K. Berggren, and D.~Englund, ``{On-chip
  detection of non-classical light by scalable integration of single-photon
  detectors},'' {\em Nature Communications}, vol.~6, p.~5873, 1 2015.

\bibitem{Schelew2015}
E.~Schelew, M.~K. Akhlaghi, and J.~F. Young, ``{Waveguide integrated
  superconducting single-photon detectors implemented as near-perfect absorbers
  of coherent radiation},'' {\em Nature Communications}, vol.~6, pp.~1--8,
  2015.

\bibitem{McCaughan2014}
A.~N. McCaughan and K.~K. Berggren, ``{A superconducting-nanowire
  three-terminal electrothermal device.},'' {\em Nano letters}, vol.~14,
  no.~10, pp.~5748--53, 2014.

\bibitem{Gehl2017}
M.~Gehl, C.~Long, D.~Trotter, A.~Starbuck, A.~Pomerene, J.~B. Wright,
  S.~Melgaard, J.~Siirola, A.~L. Lentine, and C.~DeRose, ``{Operation of
  high-speed silicon photonic micro-disk modulators at cryogenic
  temperatures},'' {\em Optica}, vol.~4, no.~3, p.~374, 2017.

\bibitem{Khurgin2015}
J.~B. Khurgin, G.~Sun, W.~T. Chen, W.-Y. Tsai, and D.~P. Tsai, ``{Ultrafast
  Thermal Nonlinearity.},'' {\em Scientific reports}, vol.~5, no.~December,
  p.~17899, 2015.

\bibitem{Vernon2017a}
Z.~Vernon, M.~Menotti, C.~C. Tison, J.~A. Steidle, M.~L. Fanto, P.~M. Thomas,
  S.~F. Preble, A.~M. Smith, P.~M. Alsing, M.~Liscidini, and J.~E. Sipe,
  ``{Truly unentangled photon pairs without spectral filtering},'' {\em arXiv},
  no.~March, pp.~1--5, 2017.

\bibitem{Tison2017}
C.~C. Tison, J.~A. Steidle, M.~L. Fanto, Z.~Wang, N.~A. Mogent, A.~Rizzo, S.~F.
  Preble, and P.~M. Alsing, ``{The Path to Increasing the Coincidence
  Efficiency of Integrated Photon Sources},'' {\em arXiv}, pp.~1--6, 2017.

\bibitem{Gimeno-segovia2017}
M.~Gimeno-Segovia, H.~Cable, G.~J. Mendoza, P.~Shadbolt, J.~W. Silverstone,
  J.~Carolan, M.~G. Thompson, J.~L. O'Brien, and T.~Rudolph, ``{Relative
  multiplexing for minimising switching in linear-optical quantum computing},''
  {\em New Journal of Physics}, vol.~19, no.~6, 2017.

\end{thebibliography}

\end{document}